\documentclass[namedreferences]{solarphysics}

\usepackage[hyperref,optionalrh]{spr-sola-addons} 
\usepackage[all]{hypcap} 
\usepackage{graphicx}        
\usepackage{amssymb}        
\usepackage{textcmds}        
\usepackage{color}           
\usepackage{breakurl}        

\usepackage{amsopn}
\usepackage{multirow}
\usepackage{gensymb}

\usepackage{psfrag}
\usepackage{auto-pst-pdf}


\usepackage{setspace}
\hypersetup{
	colorlinks = true, 
	allcolors = blue,
}

\makeatletter
\newcommand\tsb[1]{\@textsubscript{\selectfont#1}}
\def\@textsubscript#1{{\m@th\ensuremath{_{\mbox{\fontsize\sf@size\z@#1}}}}}
\newcommand\tsp[1]{\@textsuperscript{\selectfont#1}}
\def\@textsuperscript#1{{\m@th\ensuremath{^{\mbox{\fontsize\sf@size\z@#1}}}}}

\providecommand{\e}[1]{\ensuremath{\times 10^{#1}}}
\providecommand{\lsr}[1]{\textit{V}\tsb{LSR}}

\providecommand{\rsolar}[1]{#1 R\tsb{$\odot$}}

 \renewcommand*{\thetabnote}{\alph{tabnote}}

\providecommand{\edit}[1]{{\color{black}{#1}}}
\providecommand{\redit}[1]{{\color{black}{#1}}}

\DeclareMathOperator*{\argminA}{arg\,min}

\usepackage{enumitem}
\setitemize{label=\textbullet, leftmargin=*, nolistsep}

\begin{document}


\psfrag{X (rsun)}[][][0.65]{{\fontfamily{phv}\selectfont X (R$_{\odot}$)}}
\psfrag{Y (rsun)}[][][0.65]{{\fontfamily{phv}\selectfont Y (R$_{\odot}$)}}
\psfrag{Distance from Sun-Center (rsun)}[][][0.6]{{\fontfamily{phv}\selectfont Distance from Sun-Center (R$_{\odot}$)}}
\psfrag{Distance along slit (rsun)}[][][0.7]{{\fontfamily{phv}\selectfont Distance along slit (R$_{\odot}$)}}

\psfrag{X Distance from Sun-center (rsun)}[][][0.7]{{\fontfamily{phv}\selectfont X Distance from Sun-center (R$_{\odot}$)}}

\begin{article}

\begin{opening}

\title{The Low-Frequency Solar Corona in Circular Polarization}

\author[addressref={af1},corref,email={patrick.mccauley@sydney.edu.au}]{\inits{P.~I.}\fnm{Patrick~I.}~\lnm{McCauley}\orcid{0000-0002-1450-7350}}
\author[addressref={af1}]{\inits{I.~H.}\fnm{Iver~H.}~\lnm{Cairns}\orcid{0000-0001-6978-9765}}
\author[addressref={af2}]{\inits{S.~M.}\fnm{Stephen~M.}~\lnm{White}\orcid{0000-0002-8574-8629}}
\author[addressref={af3}]{\inits{S.}\fnm{Surajit}~\lnm{Mondal}}
\author[addressref={af4}]{\inits{E.}\fnm{Emil}~\lnm{Lenc}\orcid{0000-0002-9994-1593}}
\author[addressref={af5}]{\inits{J.}\fnm{John}~\lnm{Morgan}}
\author[addressref={af3}]{\inits{D.}\fnm{Divya}~\lnm{Oberoi}\orcid{0000-0002-4768-9058}}

\address[id=af1]{School of Physics, University of Sydney, Sydney, NSW 2006, Australia}
\address[id=af2]{Space Vehicles Directorate, Air Force Research Laboratory, Albuquerque, NM, USA}
\address[id=af3]{National Centre for Radio Astrophysics, Ganeshkind, Pune 411007, Maharashtra, India}
\address[id=af4]{CSIRO Astronomy and Space Science, PO Box 76, Epping, NSW 1710, Australia}
\address[id=af5]{International Centre for Radio Astronomy Research, Curtin University, Perth, WA 6845, Australia}

\runningauthor{McCauley et al.}
\runningtitle{The Low-Frequency Solar Corona in Circular Polarization}


\begin{abstract}

We present spectropolarimetric imaging observations of the solar corona 
at low frequencies (80\,--\,240 MHz) using the \textit{Murchison Widefield Array} (MWA). 
These images are the first of their kind, and we introduce an algorithm to mitigate 
an instrumental artefact \edit{by} which the total intensity 
signal contaminates the polarimetric images due to calibration errors. 
We then survey the range of circular polarization 
(Stokes $V$) features detected in over 100 observing runs near solar maximum \redit{during quiescent periods}. 
\edit{First,} we detect around \redit{700} compact polarized sources across our dataset with polarization 
fractions ranging from less than 0.5 \% to nearly 100 \%. 
These sources exhibit a positive correlation between polarization fraction and 
total intensity, and we interpret them as a continuum of plasma emission noise storm 
(Type I burst) \redit{continua} sources associated with active regions. 
\edit{Second,} we report a characteristic ``bullseye" structure observed for many low-latitude coronal holes 
in which a central polarized component is surrounded by a ring of the opposite sense. 
The central component does not match the sign expected from thermal bremsstrahlung emission, and 
we speculate that propagation effects or an alternative emission mechanism may be responsible. 
\edit{Third,} we show that the large-scale polarimetric structure at our lowest frequencies is reasonably well-correlated 
with the line-of-sight (LOS) magnetic field component inferred from a global potential field source surface (PFSS) model. 
The boundaries between opposite circular polarization signs are generally aligned with polarity 
inversion lines in the model at a height roughly corresponding to that of the radio limb. 
\edit{This is not true at our highest frequencies, however, where the LOS magnetic field direction and polarization sign 
are often not straightforwardly correlated.}

\end{abstract}
\end{opening}


\section{Introduction} %
\label{introduction} %

Radio emission in a magnetized plasma is produced in one or both of two modes, the 
ordinary [$o$] and extraordinary [$x$], which are each 100 \% circularly polarized with opposite senses 
in the quasi-circular approximation generally used for the solar corona \citep{Zheleznyakov77,Melrose80}.  
The $x$-mode refers to when the electric field vector of the electromagnetic wave rotates in the same direction 
as the gyromotion of electrons around the magnetic field where the emission was generated. 
A net circular polarization arises when the two modes are \edit{received} unequally, 
which is characterized by the degree [$r_{\rm c}$] of circular polarization [Stokes $V$] relative to the total intensity [Stokes $I$].
In detail,  

\begin{equation}
	r_{\rm c}  = \frac{T_{{\rm b},x}-T_{{\rm b},o}}{T_{{\rm b},x}+T_{{\rm b},o}}~, 
\end{equation}

\noindent where $T_{{\rm b},x}$ and $T_{{\rm b},o}$ refer to the brightness temperatures of the $o$ and 
$x$ modes, respectively \citep{Dulk85}. 
The quantity $r_{\rm c}$, also labeled dcp or $V/I$, depends on the emission 
mechanism and plasma parameters, along with a number of effects such as mode coupling and refraction 
that may modulate the polarization state or separate the two modes during propagation. 
Low-frequency (meter-wave) emission from the solar corona is dominated by two mechanisms,
thermal bremsstrahlung and plasma emission (\textit{e.g.} \citealp{Dulk85,White99,Aschwanden05}). 
Other mechanisms are also important in specific contexts, such as (gyro)synchrotron emission 
in coronal mass ejections, but \edit{these} will not be discussed in detail here. 

Bremsstrahlung emission is produced by the conversion of kinetic energy into radiant energy that 
occurs when a charged particle accelerates, and  
thermal bremsstrahlung refers to 
a plasma in thermal equilibrium for which free electrons are deflected by the Coulomb fields of ions and atomic nuclei.
This is often referred to as free-free radiation for a fully-ionized plasma like the corona because 
the particles are not in bound states throughout the entire process. 
Emission \edit{at} a particular frequency is generated only by plasma with electron densities [$n_{\rm e}$] equal to 
or below that corresponding to the \edit{local} fundamental electron plasma frequency 
[$f_{\rm p} \approx 9\e{-3}\sqrt{n_{\rm e}}$~MHz, for $n_{\rm e}$ in cm\tsp{-3}]. 
Lower-frequency emission therefore corresponds to lower-density material at 
generally larger heights above the surface, meaning that the corona appears 
larger with decreasing frequency. 
\edit{Canonical coronal background density models (\textit{e.g.} \citealp{Newkirk61,Saito77}) correspond to frequencies of 
below $\approx$ 300 MHz, but dense coronal structures may produce free-free emission well into the GHz range.}

Thermal bremsstrahlung slightly favors the $x$-mode to a degree that depends primarily on the 
line-of-sight (LOS) magnetic field strength. 
The opacity, $\kappa$, can be written as

\begin{equation} 
	\label{eq:brem}
	\kappa = 0.2\frac{n_{\rm e}^2}{T_{\rm e}^{1.5}(f \pm f_B|\cos{\theta}|)^2}~{\rm cm}^{-1}, 
\end{equation}

\noindent where $T_{\rm e}$ is the electron temperature, $f$ is the 
emission frequency, $f_B$ is the electron gyrofrequency [$f_B = 2.8\e{6}B_{\rm gauss}$ Hz], and 
$\theta$ is the angle between the line of sight and the magnetic field direction \citep{Dulk85,Gelfreikh04,Gibson16}. 
The plus sign refers to the $o$-mode, the minus sign refers to the $x$-mode, and the difference 
between the two modes produces the net circular polarization. 
\redit{Equation~\ref{eq:brem} is a quasi-linear (QL) approximation that is valid for most angles $\theta$. 
Values of $\theta$ close to 90$\degree$, for which the propagation direction is nearly perpendicular to the magnetic field 
orientation, are referred to as quasi-transverse (QT) propagation and produce linear polarizations \citep{Zheleznyakov70,Ryabov04}. 
Circularly-polarized emission that passes through a QT region may also experience polarization state changes, which will be 
discussed in Section~\ref{discussion}}.
Equation~\ref{eq:brem} also assumes that $f \gg f_B$. 
This condition means that the difference between the two modes, and therefore the polarization 
fraction, will always be fairly small, generally a few percent or less \edit{at the low frequencies considered in this article} \citep{Sastry09}. 
For a homogenous, optically-thin plasma, $r_c \approx 2 \cos{\theta}(f_B/f)$, \edit{while}  
for the optically-thick case, a temperature gradient is required for the two modes to be produced unequally \edit{\citep{Dulk85,Gibson16}}.  

Thermal bremsstrahlung radiation generates a continuous background  
that slowly varies as the corona evolves. 
This may be slightly or dramatically augmented by transient emission associated with 
nonthermal electrons that are accelerated through a variety of mechanisms underpinned 
either by magnetic reconnection or shock waves. 
These electron streams produce oscillations in the background plasma known as 
Langmuir waves, which then deposit energy into radio emission through scattering 
by ion sound waves or by other nonlinear Langmuir wave processes \citep{Ginzburg58,Robinson00,Melrose09}. 
These are typically coherent mechanisms, often grouped together under the term ``plasma emission," for which the 
intensity is related nonlinearly to the energy of the nonthermal electrons. 
Plasma emission is responsible for most types of solar radio bursts \citep{Dulk85}, which may 
exceed the thermal background by several orders of magnitude, but it is also likely 
the source of very weak nonthermal emissions that enhance the background only slightly \citep{Suresh17,Sharma18}. 
Like thermal bremsstrahlung, plasma emission is tied to the ambient 
density though the electron plasma frequency. 
However, in this case, the emission frequency is highly localized to just above the plasma frequency or 
its harmonic.  

The polarization of plasma emission depends firstly on the harmonic number. 
For fundamental [$f_{\rm p}$] emission, the circular polarization fraction should be 100 \% in the sense of 
the $o$-mode because\edit{, for frequencies expressed in Hz,} $f_{\rm p}$ is above the cutoff for $x$-mode production, meaning that $x$-mode radiation  
begins only at frequencies slightly lower than the plasma frequency \citep{Melrose09}. 
Polarization fractions approaching 100 \% are indeed sometimes observed for Type I bursts (\textit{e.g.} \redit{\citealp{Kai62,Tsuchiya63,Dulk84,Aschwanden86,Mugundhan18}}). 
\edit{However,} this is almost never true for other radio burst types that are also attributed to fundamental plasma emission (\textit{e.g.} \citealp{Wentzel84,Reid14,Kaneda15}). 
\edit{The} reason for this remains an open question, but a common explanation is that scattering of the 
radio emission by other wave modes or by sharp density gradients tends to have a depolarizing effect (\textit{e.g.} \citealp{Wentzel86,Melrose89,Melrose06,Kaneda17}). 
The polarization fraction of harmonic [$2f_{\rm p}$] emission is more complicated because it depends on the 
angular distribution of the Langmuir waves. 
Polarization in the sense of the $o$-mode is still generally expected, assuming that the Langmuir waves are confined to relatively small 
angles with respect to the magnetic field, which is generally assumed to be true because of the associated magnetic field strengths \citep{Melrose78}. 
However, it is possible for the $x$-mode to dominate in specific\edit{, and likely less common,} contexts \citep{Willes97}. 
Thus, for the same LOS magnetic field direction, the two dominant low-frequency emission mechanisms generally produce opposite 
circular polarization signatures. 
 
Radio polarimetry has long been a powerful tool for diagnosing solar magnetic fields, particularly using 
high-frequency observations of gyroresonance emission \citep{Akhmedov82,White97} and, more recently, 
bremsstrahlung emission \citep{Grebinskij00}.
Low-frequency polarimetry has generally been restricted to radio bursts  
because their high intensities and large polarization fractions are easiest to detect. 
An early review on the polarization of metric bursts and their utility as magnetic field probes is given by \citet{Dulk78}.
Very few instruments have been capable of making two-dimensional polarimetric measurements of the low-frequency Sun, 
and until now, none have been sensitive enough to detect the weak polarization signatures during quiescent periods. 
In recent decades, this type of analysis could be done with two instruments, the Nan{\c c}ay Radioheliograph (NRH; \citealp{Kerdraon97}), 
which operates between 150 and 450 MHz, 
and the Gauribidanur Radioheliograph (GRH; \citealp{Ramesh98}), which \edit{usually} operates at 80 MHz. 

A few studies have ulilized the polarimetric imaging capabilities of the NRH to examine spatial variation in radio bursts. 
For example, \citet{Mercier90} showed that Type III bursts have different spatial characteristics 
in circular polarization compared to the total intensity, and 
\citet{Bouratzis16} investigated similar differences in spike bursts as a function of time. 
Several others have examined source positions and structures in total intensity NRH observations, while using the polarization 
information to help discriminate between emission mechanisms (\textit{e.g.} \citealp{Gopalswamy94,Tun13,Kong16,Liu18}).
The radioheliograph at Gauribidanur does not have a polarimetric capability itself, but several one-dimensional 
polarimeters have been installed alongside it \citep{Ramesh08,Sasikumar13,Kishore15}. 

GRH imaging and simultaneous polarimeter observations have been used for studies of 
Type I noise storms \citep{Ramesh11,Ramesh13,Mugundhan18}, Type II bursts \citep{Hariharan14,Hariharan15,Kumari17}, 
Type III bursts \citep{Ramesh10b,Sasikumar13b,Kishore17}, Type IV bursts \citep{Hariharan16}, and 
gyrosynchrotron emission from CMEs \citep{Sasikumar14}. 
Most of these results include estimates of the associated magnetic field strength assuming a particular emission mechanism. 
Additionally, \citet{Ramesh10} report polarized emission from streamers that is attributed to thermal bremsstrahlung, though 
the polarization fraction ($\approx$ 15 \%) is unusually large for bremsstrahlung emission. 
Moreover, the polarized source cannot be localized beyond assuming that it comes from the dominant total intensity source, and 
as we will show, polarized emission from the low-frequency corona is often not straightforwardly correlated with total intensity, particularly 
during quiescent periods. 

This article presents the first spectropolarimetric imaging observations \edit{of the Sun} from the \textit{Murchison Widefield Array} (MWA; \citealp{Tingay13}). 
These are the first circular polarization images of the low-frequency corona that are sensitive enough to detect the \edit{polarimetric} signatures 
associated with thermal bremsstrahlung emission and very weak plasma emission outside of major burst periods. 
We will survey the range of features detected in over 100 observing runs near solar maximum and motivate future studies with these novel data. 
Section~\ref{mwa} describes the MWA instrument, and Section~\ref{leakage} introduces an algorithm used to mitigate 
an important calibration artefact. 
Section~\ref{ar} discusses active region noise storm sources, Section~\ref{ch} characterizes the polarimetric signature of 
coronal holes, and Section~\ref{qs} details the large-scale quiescent structure. 
We discuss the implications of our results and motivate future studies in Section~\ref{discussion}. 
Our conclusions are summarized in Section~\ref{conclusion}. 


\section{\textit{Murchison Widefield Array} (MWA)}
\label{mwa}

The MWA is a low-frequency radio interferometer \edit{located} in Western Australia \citep{Lonsdale09,Tingay13}, and  
heliophysics is among the instrument's principal science themes alongside 
astrophysical topics \citep{Bowman13}. 
Direct solar observations have characterized the weakest nonthermal 
emissions reported to-date \citep{Suresh17,Sharma18}, provided definitive evidence 
for the standard theory of Type III bursts \citep{Cairns18}, detailed new radio 
burst dynamics \citep{McCauley17,Mohan18}, used radio bursts to probe the 
coronal density structure \citep{McCauley18}, characterized the low-frequency 
signature of coronal holes \citep{Rahman19}\edit{, and provided evidence for coronal heating 
via weak particle acceleration episodes \citep{Mohan19}.  
Solar imaging with the MWA has also motivated advances in data processing techniques 
related to flux calibration \citep{Oberoi17}, spatially resolved dynamic spectra \citep{Mohan17}, and high dynamic range imaging \citep{Mondal19}.}
Additionally, widefield interplanetary scintillation observations may be used for studies 
of the solar wind and of coronal mass ejections (CMEs) propagating through the heliosphere \citep{Kaplan15,Morgan18}. 

The MWA is comprised of \edit{4096} dipole antennas 
arranged in 128 aperture arrays called ``tiles". 
This refers to the Phase I array used here, which began observing in \edit{2013}.
An expanded Phase II array began full operations in 2018 with twice as many tiles\edit{, of 
which 128 can be used simultaneously in different configurations \citep{Wayth18}.}
The MWA has an instantaneous bandwidth of 30.72 MHz that can be distributed 
between 80 and 300 MHz in various configurations.
Our data utilize a ``picket fence" mode with 12 contiguous 2.56 MHz bandwidths 
centered at 80, 89, 98, 108, 120, 132, 145, 161, 179, 196, 217, and 240 MHz.
The data were recorded with a 0.5 sec time \edit{resolution} and a 40 kHz spectral resolution, 
but the observations presented here are averaged over each 2.56 MHz bandwidth before 
imaging and then time-averaged to different degrees after imaging. 
\redit{The spatial resolution is defined by the synthesized beam sizes, which have 
major axes of around 6.4 arcmin (\rsolar{0.40}) at 80 MHz and 2.5 arcmin (\rsolar{0.16}) at 240 MHz. 
The beam sizes and orientations, shown in the lower-left corners of each image, vary somewhat 
between observations due to pointing differences and occasional antenna failures.}

We use the same data processing scheme as \citet{McCauley17} and \citet{McCauley18}, and 
what follows is a brief summary thereof. 
Visibilities were generated with the standard MWA correlator \citep{Ord15} and the  
\textsf{cotter} software \citep{Offringa12,Offringa15}. 
Observations of bright and well-modelled calibrator sources were used 
to obtain solutions for the complex antenna gains \citep{Hurley14}, which 
were improved by imaging the calibrator and iteratively self-calibrating from there \citep{Hurley17}. 
All of our observations were calibrated using either Centaurus A or Hercules A. 
\textsf{WSClean} \citep{Offringa14} was used to perform the imaging 
with a Briggs -2 weighting \citep{Briggs95} to emphasize spatial resolution and minimize point spread function (PSF) sidelobes. 
The primary beam model of \citet{Sutinjo15} was used to produce Stokes $I$ and $V$
images from the instrumental polarizations, 
and the SolarSoftWare (SSW; \citealp{Freeland98}) 
routine \textsf{mwa\_prep} \citep{McCauley17} was used to translate the images onto solar coordinates. 
\redit{The data presented here are not flux calibrated on an absolute scale. 
Intensities are expressed either relative to the Stokes $I$ background level or in units of signal-to-noise.}

The next section will describe further steps required to calibrate the polarization images. 
\edit{We use the International Astronomical Union (IAU) and Institute of Electrical and Electronics Engineers (IEEE) convention on circular polarization, which defines 
positive as being right-handed (clockwise) from the source's perspective \citep{IEEE69,IAU73}, 
where right-handed refers to the rotation of the electric field vector of the electromagnetic wave about the 
orthogonal direction of motion. 
This convention is convenient here because it means that a net polarization in the sense of the $x$-mode 
will match the sign of the line-of-sight magnetic field component [\redit{$B_{\rm LOS}$}], 
where positive is outward.}

Each observation period lasted around 5 minutes, and a total of 111 such periods in 2014 and 2015 were reduced. 
52 of these were imaged at the full 0.5-sec time resolution and 59 \edit{were sampled} at a 4-sec cadence. 
Our objective is to survey the longer-lived features that 
are present in the corona on timescales of at least minutes, outside of transient radio burst periods.  
All of the images presented in this article are median averages of the individual 0.5-sec integrations 
\edit{with total intensities that are within two standard deviations of the background level during} each 5-min observing window.
\edit{Depending on the sources present, these averaged background images} may still contain significant nonthermal emission. 
Identifying which images to include in the average is done automatically using the 
baseline procedure illustrated by Figure 5 of \citet{McCauley18}. 
This involves finding the total intensity in each image, excluding times for which the intensity 
is greater than two standard deviations above the median, and iterating until no more images 
are excluded. 
Each pixel in the output image then contains the median of the corresponding pixels 
in those low-intensity images. 
The consideration of time dependent behavior on scales of less than 5 minutes will be a topic 
of future work.


\section{\edit{An Algorithm to Mitigate the Leakage of Stokes $I$ into $V$}}
\label{leakage}

To obtain useful polarimetric images, it is necessary to account for possible ``leakage" of the Stokes $I$ signal into 
the other Stokes \edit{parameters}. 
\redit{The MWA uses dual-polarization dipole antennas arranged in 4$\times$4 grids, or ``tiles", where the 
signals for each tile component are combined in an analog beamformer that produces two outputs representing orthogonal 
$X$ and $Y$ linear polarizations \citep{Tingay13}. 
The beamformer outputs are correlated into products that fully describe 
the polarization state in ``instrumental" polarizations ($XX$, $YY$, $XY$, and $YX$) 
that may be converted into the standard Stokes parameters ($I$, $Q$, $U$, and $V$) using a model of the MWA beam pattern. 
However, there are significant differences between the analytic beam pattern and that measured empirically 
by imaging known sources \citep{Sutinjo15}. 
These differences between the actual instrumental response and the complex primary beam model lead to 
``leakage" errors in the Stokes images where some fraction of $I$ contaminates the other parameters.}

A \redit{more} detailed description of this problem and of MWA polarimetry in general is given by \citet{Lenc17}. 
\redit{Sources of discrepancy between the beam model and the true response include} imperfections in the model itself along with instrumental effects, such as 
individual dipole failures during a particular observing run, that may cause the true response to 
vary from an otherwise perfect beam model. 
\edit{Importantly, the polarimetric response is also affected by a source's position within the beam 
and zenith angle, which means that the response changes somewhat between the observation used to calibrate 
the array and the solar observation. 
Changes in the ionosphere over the 
$\lesssim$ \edit{5} hours between the calibrator and solar observations may also degrade the calibration solution.}
Our data were reduced using the \citet{Sutinjo15} beam model, which dramatically reduced leakage 
from Stokes $I$ into Stokes $Q$ but somewhat increased the leakage from $I$ into $V$ compared to previous beam models. 

The leakage fraction also varies with a source's position on the sky and its position within the field-of-view \edit{for a given calibration solution} \citep{Sutinjo15,Lenc17,Lenc18}. 
Sources observed at lower elevations and/or near the edge of the field tend to exhibit higher leakage fractions. 
It is possible to reduce the leakage by means of iterative self-calibration on the source of interest, but 
this may affect the polarimetric calibration in ways that are difficult to understand, and self-calibration can also be 
difficult to effectively apply to diffuse sources like the Sun. 
Instead, the leakage effect can be mitigated with an empirical correction if there are sources within the field for which 
the polarization fractions are known. 
For the very large fields typical of many astrophysical MWA observations, a two-dimensional fit to the leakage 
fraction may be obtained from the known sources scattered throughout the field. 
The leakage fraction may vary by as much as 8 \% across a 25 deg$^2$ patch \citep{Lenc17}, 
but we do not expect significant variations across the spatial extent of the radio Sun ($\lesssim{}$1.2$^{\degree}$ at 80 MHz) 
or over the duration of a typical observation ($\approx$ 5 min). 

For the solar observations presented here, it is not possible to simultaneously observe background sources alongside the Sun due 
to limited dynamic range. 
Recent advances in calibration techniques may enable this capability \citep{Mondal19}, but those methods
\edit{cannot yet be used for polarimetry and are not used here.}
In other words, the astronomical sources that may be present are too faint to be observed in close proximity to the Sun 
and cannot be used to characterize the Stokes $I$ into $V$ leakage.
We also do not know what the polarization fraction of any particular region on the Sun should be at any given time, as 
the polarization fraction may vary considerably depending on the dominant emission mechanism and local plasma parameters. 
However, outside of radio bursts, solar emission at low frequencies is dominated by the thermal bremsstrahlung (free-free) 
process. The importance of this is that under normal quiet-Sun conditions at MWA wavelengths, 
we can expect to see bremsstrahlung radiation in most locations that is only 
slightly polarized, and we can use this statistical information to estimate the leakage fraction with an algorithm that minimizes 
the number of pixels with polarization fractions greater than some threshold (\textit{\textit{i.e.}} $|V/I| > r_{\rm c, thresh}$). 

To determine this threshold, we generated synthetic Stokes $V/I$ images for each of our observing periods and frequencies using 
the forward modeling code \textsf{FORWARD} \citep{Gibson16} in SolarSoft IDL. 
\edit{FORWARD calculates the Stokes $I$ and $V$ intensities expected from thermal bremsstrahlung emission using Equation~\ref{eq:brem}, 
with the temperature, density, and magnetic field parameters taken in this case from the month-averaged 
Magnetohydrodynamic Algorithm outside a Sphere (MAS; \citealp{Lionello09}) global coronal model.}
On average 50 \% of pixels in these images with Stokes $I$ brightness temperatures greater than 100,000 K have fractional polarizations of 
less than 0.3 \%.
We choose a slightly larger threshold of 0.5 \% because we wanted to implement our procedure uniformly, and 
the noise level in some of our observations makes a lower threshold impractical.
This value is also consistent with the predictions of \citet{Sastry09}, and the
effect of varying the threshold is folded into $V/I$ uncertainty estimates presented in Section~\ref{ch}. 

Our algorithm therefore assumes that most of the pixels in our images of the quiescent corona should exhibit polarization 
fractions of less \edit{than} 0.5 \% and determines the leakage fraction that minimizes the number of pixels with $V/I$ values 
greater than 0.005. The algorithm can be expressed formally as:

\begin{equation}
\label{eq:alg1}
f(L) = \sum_{k=1}^{n} \left[ \left| \frac{V_k - L\cdot{}I_k}{I_k} \right| > r_{\rm c, thresh} \right]
\end{equation}

\begin{equation}
\label{eq:alg}
L_{\rm min} = \argminA_{L\in(-1,1)} f(L),
\end{equation}

\noindent \edit{where $f(L)$ is the number of pixels with polarization fractions greater than $r_{\rm c, thresh}$ as a function 
of $L$, the constant fraction of Stokes $I$ that is assumed to have leaked into Stokes $V$.}
The aim is to find the value \edit{$L_{\rm min}$} that minimizes \edit{$f(L)$}, where $k$ is a given pixel in an 
image and $n$ is the number of pixels to be considered. 
We consider only pixels for which a Stokes $I$ signal is detected above 5 $\sigma$. 
The square brackets \edit{in Equation~\ref{eq:alg1}} refer to the Iverson bracket notation, meaning that their contents evaluate to 1 if 
the condition is satisfied and 0 otherwise. 
In this case, that simply means that a pixel is counted if its polarization fraction is greater than 0.005 (0.5 \%). 
Equation~\ref{eq:alg} is evaluated using an adaptive grid search with increments in $L$ of 0.1, 0.01, and 0.001.  
Note that this strategy is not the same as minimizing the total polarized intensity, which is not advisable because the two senses 
may not be equally represented and specific regions may have large polarized intensities that would bias the result if 
one were to simply find $L$ that minimizes the total polarization fraction in the image. 


\begin{figure*}
    \centering
    \includegraphics[width=1.0\textwidth]{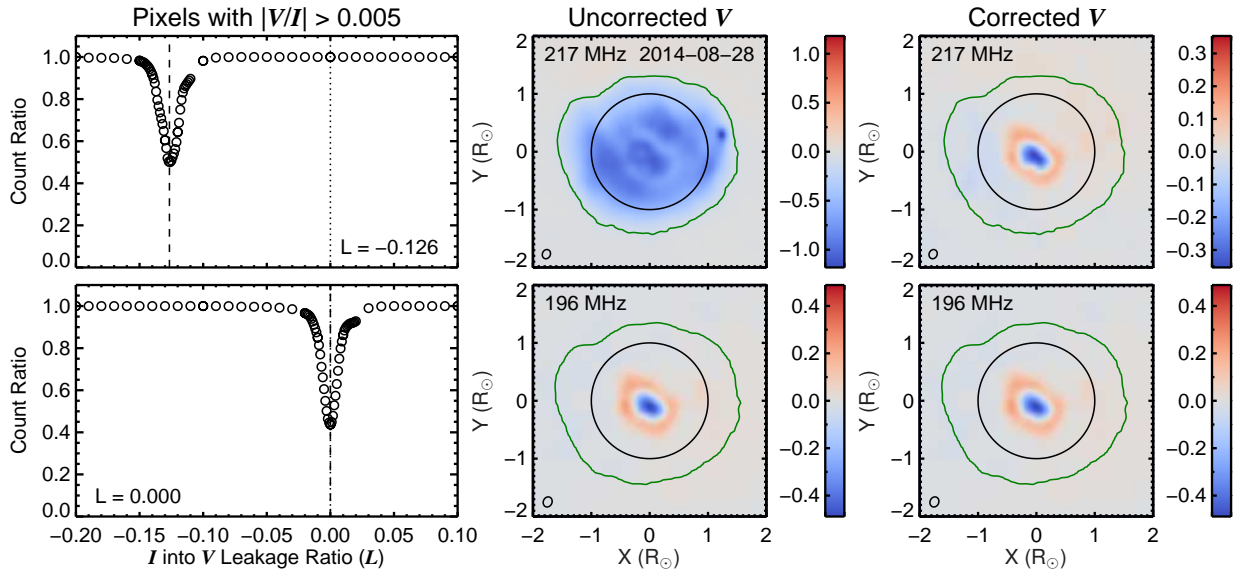}
    \caption{\footnotesize{
	An illustration of the leakage subtraction algorithm. The left panels show the implementation of 
	Equation~\ref{eq:alg} to find the $L$ that minimizes the number of pixels with $|V/I| > 0.005$. 
	At 217 MHz (top), we estimate that -12.5 \% of the Stokes $I$ signal leaked into Stokes $V$, and 
	at 196 MHz, the same procedure estimates there to be no leakage. The middle panels show 
	the uncorrected Stokes $V$ images, and the right panels show the corrected images 
	(\textit{\textit{i.e.}} $V - L\cdot{}I$). \redit{The corrected image at 217 MHz is shown as a function of $L$  
	in an animated version of this figure available in the \href{http://www.physics.usyd.edu.au/~pmcc8541/mwa/polarization/index.html}{online material}.} 
       }}
    \label{fig:lsub1}
\end{figure*}


\begin{figure*}
    \centering
    \includegraphics[width=1.0\textwidth]{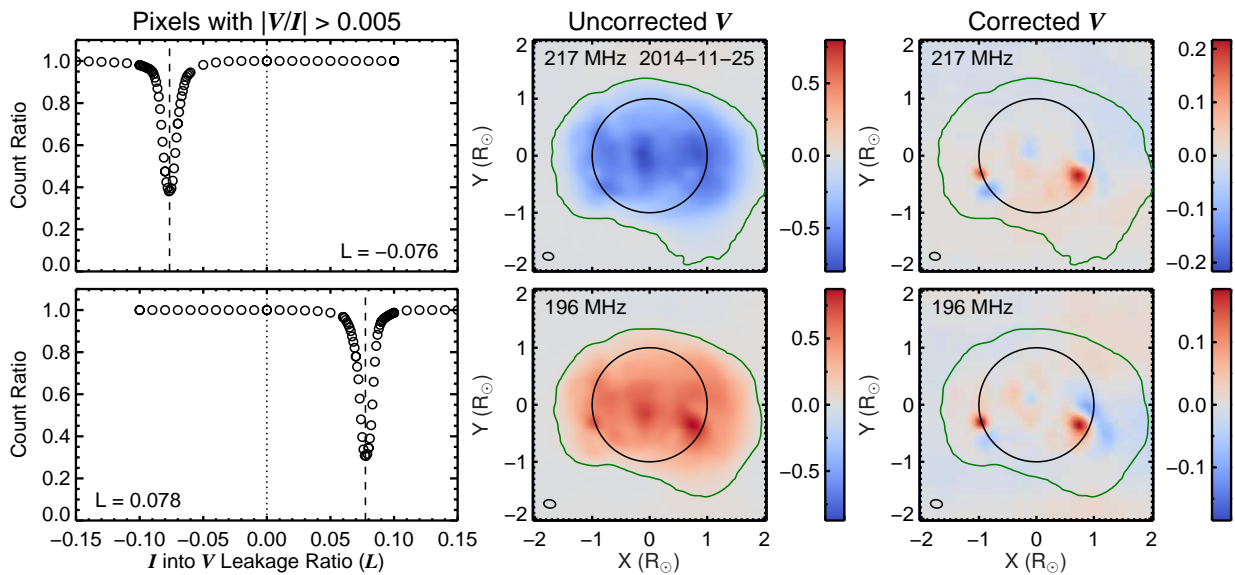}
    \caption{\footnotesize{
	Same as Figure~\ref{fig:lsub1} but for a different observation with different structures and a different leakage 
	behavior. 
       }}
    \label{fig:lsub2}
\end{figure*}

Figures~\ref{fig:lsub1} and \ref{fig:lsub2} show the result of applying this algorithm to images from two frequency channels 
on two different days. In Figure~\ref{fig:lsub1}, we see that the uncorrected Stokes $V$ image at 217 MHz \edit{would imply} that the 
entire corona is highly circularly polarized with a single sense, while the uncorrected 196 MHz image suggests a very 
different structure with a mixture of opposite signs. 
We know that entire corona at 217 MHz should not be polarized with a single sense to the extent implied by the uncorrected 
217 MHz image. 
Our algorithm suggests that $L$ = -0.125 in this case, and applying that correction recovers the same structure that is apparent in the 
196 MHz image. 
Importantly, the same procedure can also be applied to observations for which there is little or no leakage, as is illustrated by the 
196 MHz example in Figure~\ref{fig:lsub1}. 
Figure~\ref{fig:lsub2} shows another example for which the two frequency channels shown are impacted by significant leakage of opposite signs, but 
once corrected, they exhibit very similar structures.  


\begin{figure*}
    \centering
    \includegraphics[width=1.0\textwidth]{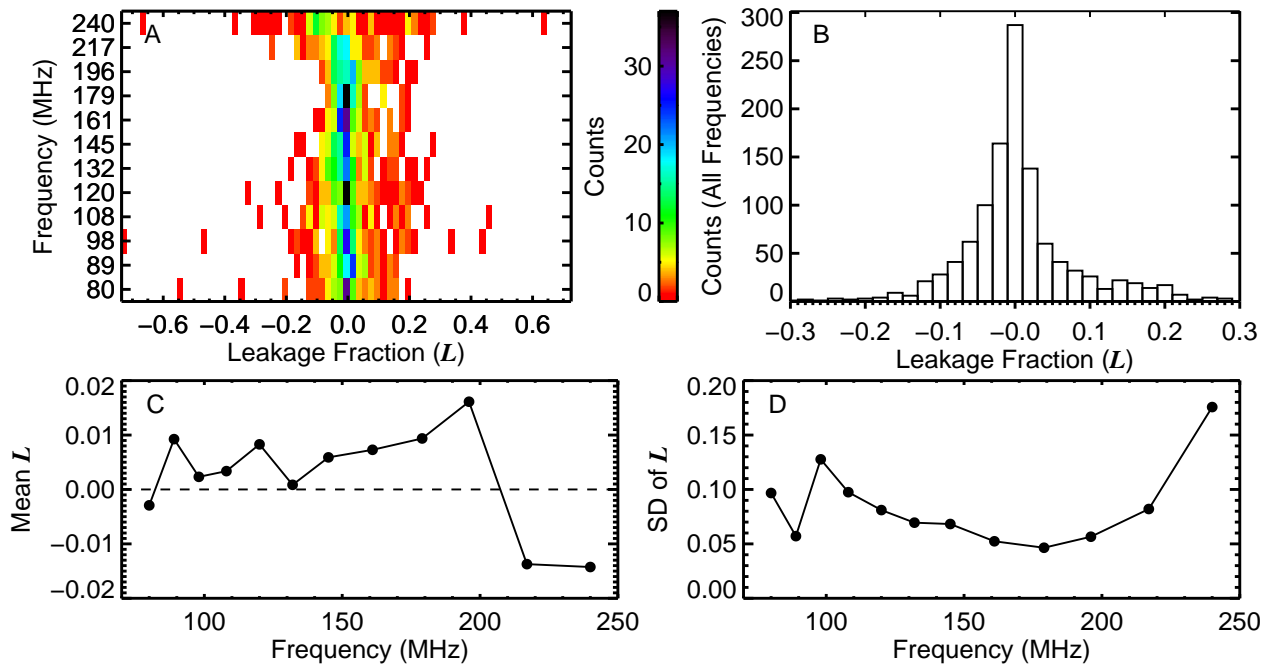}
    \caption{\footnotesize{
    	Summary of results from applying the Equation~\ref{eq:alg} algorithm to 1144 different observations. 
	\textbf{(A):} Two-dimensional histogram showing the leakage fraction [$L$] as a function of frequency with a 
	bin size of 0.02 in $L$. 
	\textbf{(B):} Histogram of $L$ across all frequencies.  
	\textbf{(C):} The average $L$ for each frequency channel. 
	\textbf{(D):} The standard deviation of $L$ for each frequency channel. 
	Around 26 \% of observations have $|L| < 0.01$, and around 84 \% have $|L| < 0.1$. 
	Panels A and D show that the leakage is most severe and variable at the extremes of 
	the bandwidth.
       }}
    \label{fig:lhist}
\end{figure*}

The examples in Figures~\ref{fig:lsub1} and \ref{fig:lsub2} are cases for which the leakage is fairly severe. 
Figure~\ref{fig:lhist} summarizes the leakage behavior across our dataset.
\redit{83 \% of the images exhibit leakage fractions less than or equal to 10 \%. 
$L \leq$ 0.05 and 0.01 for 66 \% and 25 \% of the images, respectively.} 
Figure~\ref{fig:lhist}d shows that the standard deviation of $L$ is lowest at 179 MHz, which is consistent with the 
astrophysical MWA literature that indicates the leakage tends to be worst near the ends of the bandwidth. 
This summary includes 106 different sets of spectroscopic imaging observations for a total of 1144 images to which the algorithm could be applied.  
In Section~\ref{mwa}, we stated that 111 observing periods were analyzed, which would imply 1332 images given our 12 frequency 
channels.
Some images are rejected for polarimetry because they do not contain enough pixels detected above 5 $\sigma$ in Stokes $I$, 
generally because a nonthermal active region source is so intense as to elevate the noise floor above the level of the thermal 
disk. 
\redit{In other words, there is insufficient dynamic range to simultaneously detect both the thermal and 
nonthermal components present at those times.}

\redit{These observations must be excluded because the algorithm relies on the statistical expectation that most pixels are 
dominated by thermal bremsstrahlung emission, and in} these cases, there are not enough ``thermal" pixels with 
sufficient signal-to-noise ratios for the algorithm to function.
\redit{Images were rejected for polarimetry if they contained fewer pixels above 5 $\sigma$ in 
Stokes $I$ than that enclosed by a circle of radius equal to the height of the plasma frequency layer at a given frequency given a 
3$\times$ \citet{Newkirk61} density model, which roughly approximates the height of the radio limb in our observations.
Around 14 \% of our data failed this test and are excluded from further analysis.}
\redit{It is important to note that this} introduces a bias in the next section on active region sources because the most intense sources tend to be the most 
highly polarized, but the leakage artefact cannot be constrained using our method for the brightest among them. 
The intense and highly-polarized population is therefore very likely to be underrepresented. 

\redit{Leakage may also occur from Stokes $U$ into $V$ \citep{Lenc17}, but this is not a concern here because linear polarizations from the 
corona are negligible at our frequencies and observing bandwidths (\textit{e.g.} \citealp{Gibson16}).
Of potential concern, however, is possible leakage from $V$ into $I$. 
This could add to or subtract from the Stokes $I$ levels, decreasing the reliability of measured polarization fractions. 
Unlike leakage from Stokes $I$ into the other parameters, 
the reverse case has not been investigated for the MWA because the polarization fractions of astrophysical sources 
are generally so low as to make this effect very difficult to characterize and unlikely to significantly impact the results. 
However, solar radio bursts may have large circular polarization fractions, meaning that Stokes $V$ into $I$ leakage 
could be a significant contaminant in some cases. 
We currently have no way to assess or mitigate this contamination, but 
we anticipate that the effect should occur at a similar or lower level than leakage from Stokes $I$ into $V$, 
as the mechanism would be similar but with generally lower magnitudes.}

\redit{Of concern are sources with high polarization fractions and large leakage fractions. 
This is relevant mainly for the next section, which focuses on nonthermal active region sources. 
Assuming that $V$ into $I$ leakage may occur at up to the same level estimated for $I$ into $V$, 
this introduces an uncertainty in $V/I$ of less than 1 \% for 79 \% of the sources 
and an uncertainty of less than 5 \% for 95 \% of the sample described by Figure~\ref{fig:ars_hist}. 
The remainder have uncertainties of 10 \% on average, and up to 19 \% for one event, due to this effect. 
We have chosen to represent this latter population with a different symbol in Figure~\ref{fig:ars_hist} 
to indicate that their polarization fractions should be treated with additional skepticism.
For Figure~\ref{fig:ch_slices} in Section~\ref{ch}, we estimate uncertainties on $V$/$I$ of $\lesssim$ 3 \% 
by combining several effects for two cases where different observations on the same day with different values of $L$ could be compared. 
These observations correspond to thermal or very weak nonthermal emission for which the polarization 
fractions are lower than 5 \%.
Potential $V$ into $I$ leakage is therefore not a significant concern in that case 
and constitutes an average of 14 \% of the total 
error bars in Figure~\ref{fig:ch_slices}.}


\section{Active Region Noise Storm Sources}
\label{ar}

The most common features in these images are compact polarized sources, 
the most intense of which are identified here as \redit{noise storm continua associated 
with Type I bursts.}
This is apparent from the variability in their associated dynamic spectra along with their high polarization fractions. 
However, as we will see, there are also very weak and weakly-polarized sources for 
which the source type and emission mechanism is less obvious. 

Noise storms are periods of extended burstiness that are associated with active regions and may 
persist for several days as an active region transits the disk. 
They are characterized by many distinct, narrowband \redit{Type I} bursts, often with enhanced continuum 
emission around the same frequency range \citep{Elgar77,Klein98}. 
As our data reflect the background levels during each observation period, our detections correspond to 
the continuum enhancement, along with any burst periods that could not be filtered out by our baseline 
procedure because they occurred on timescales less than the 0.5-sec time resolution. 
Despite decades of study, there are a number of unanswered questions about the  
nature of Type I bursts. 
Not all active regions that are productive at other wavelengths produce noise storms, and 
the non-radio signatures are often scant \citep{Willson05,Iwai12,Li17}, unlike Type II and III bursts, which have obvious 
associations with CMEs and flares \citep{Cairns03,Reid14}. 
There is general agreement that \redit{both the burst and continuum components of} noise storms are produced by plasma emission, largely due 
to their often high circular polarizations \citep{Aschwanden86,Mugundhan18}, but what accelerates the electrons is still debated.
Small-scale reconnection events \citep{Benz81} or weak shocks associated with upward-propagating waves \citep{Spicer82} are the 
two leading ideas, and recent work has favored persistent interchange reconnection between open and closed 
fields at the boundaries of active regions \citep{Del11,Mandrini15} or reconnection driven by moving magnetic features \citep{Bentley00,Li17}.

To automatically detect these features in the Stokes $V$ images, we developed a simple 
algorithm that begins \edit{with suppressing any diffuse polarized emission that may be present 
by applying a Butterworth bandpass filter to the fast Fourier transform (FFT) of each image. 
 \redit{The filter aims to flatten the frequency response over a particular passband, in this case the FFT frequencies corresponding to larger spatial scales, without 
 producing sharp discontinuities between the filtered and unfiltered frequencies \citep{Butterworth30}.} 
The filtered FFT is transformed back, and the resulting} image is thresholded into two binary masks, one for each polarization 
sense, that include pixels with values above the larger of 10 $\sigma$ or 20 \% of the maximum value. 
Ellipses are fit to all of the contiguous regions in the masks, and several criteria are imposed to 
obtain the final detections. 
These criteria include ensuring that 
1) the signal-to-noise ratio of pixels pulled from the 
filtered image are above 10~$\sigma$ in the original image and are of the same polarization sense,
2) the areas of the fitted ellipses are within 0.75\,--\,1.5$\times$ that of the corresponding  
synthesized beam for a given frequency, 
3) the fitted ellipses have aspect ratios no more than 1.1$\times$ that of the \edit{synthesized} beams, 
and 4) the masked regions are sufficiently elliptical, which we defined as filling at least 95 \% of the fitted ellipse. 


\begin{figure*}
    \centering
    \includegraphics[width=1.0\textwidth]{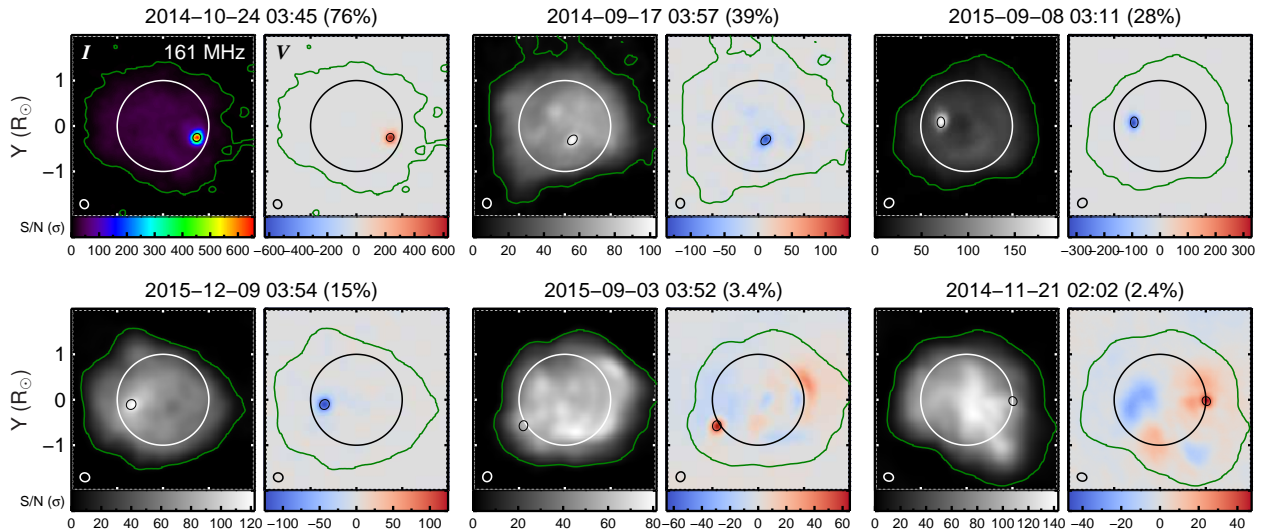}
    \caption{\footnotesize{
	Six randomly chosen examples of compact polarized sources detected at 161 MHz, sorted by 
	polarization fraction.
	The titles correspond to UTC times followed by the peak polarization fraction in parentheses.
	The color scales are linear, and the first example is plotted with a 
	different color scheme to better reflect the dynamic range of that observation. 
	Color bar intensities are expressed in units of signal-to-noise [$\sigma$], and the green 
	contour reflects the 5-$\sigma$ level in Stokes $I$. 
	Black ellipses around the sources show the region identified by the source finder algorithm. 
	The large solid circles represent the optical disk, and the ellipses in the lower-left corners 
	represent the synthesized beam sizes. 
       }}
    \label{fig:ars}
\end{figure*}

These criteria all serve to eliminate false positives that arise from the bandpass filtering, which may amplify 
noise, introduce artefacts near very bright sources, and/or not entirely suppress the large-scale diffuse emission. 
Adjusting the tolerance parameters of this algorithm can satisfactorily extract sources from any given image, 
but finding a set of defaults that could serve the entire dataset was somewhat difficult. 
We opted to aggressively tune the parameters to eliminate false positives at the cost 
of excluding false negatives. 
This procedure is run independently for all of the frequency channels in a given observation. 
The detections are then grouped across frequencies by checking for overlap among the fitted 
ellipses.
Only sources that are detected in at least three frequency channels are kept and incorporated into the 
following plots.
\edit{We find 693 sources with this method from 112 separate regions, and at least one source 
is found on 64 out of 82 days (78 \%). 
Solar Cycle 24 peaked in April 2014, and our data correspond to between August 2014 and December 2015, 
meaning that we are examining the early part of the declining phase in the solar cycle. 
As these features are associated with active regions, we would likely have found a higher fraction of days with at 
least one noise storm if our observations were shifted one year earlier and a lower fraction if the observations were taken in subsequent years.} 

Figure~\ref{fig:ars} shows six randomly-selected examples of these sources at the center 
of our bandwidth, 161 MHz. 
They exhibit polarization fractions ranging from 2.4 to 76 \% and well represent the range of sources found.
Most sources in the full sample are unipolar and are fairly isolated in the polarization images, and those with very low polarization fractions 
are sometimes embedded in diffuse emission of the same sign. 
A small number of bipolar sources were also found. 
This is somewhat inconsistent with \citet{White92}, who found that bipolar sources were nearly as common as unipolar 
sources in 327 MHz Very Large Array (VLA) observations. 
Bipolar sources are presumably less common in our observations because we are looking at lower frequencies for 
which the emission is generated at a larger height and the spatial resolution is lower. 
Preliminary analysis has revealed interesting potential anti-correlations in the intensities of the two components of 
one bipolar source, and this sort of time variability may be explored in future work. 


\begin{figure*}
    \centering
    \includegraphics[width=1.0\textwidth]{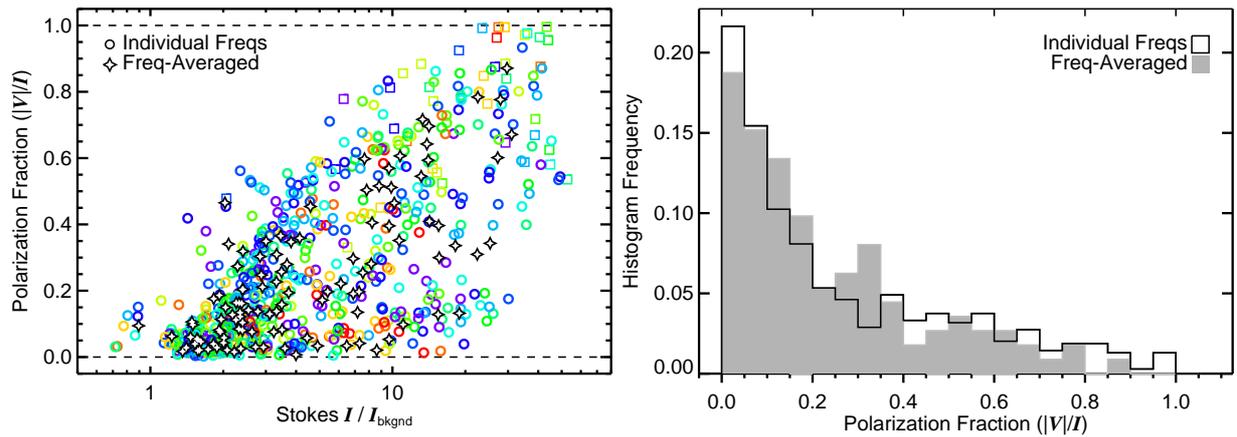}
    \caption{\footnotesize{
	\textit{Left}: Scatterplot of \redit{polarization fraction [$|V|$/$I$] versus} Stokes $I$ intensity for \edit{693}  
	compact polarized sources. 
	Values are measured at the location of peak Stokes $V$ intensity, and the Stokes $I$ intensities 
	are normalized by the median intensity of pixels detected above 5-$\sigma$ [$I_{\rm bkgnd}$]. 
	Colored circles represent measurements from individual frequency channels; purple refers to the 
	highest frequency (240 MHz) and red refers to the lowest (80 MHz). 
	All sources are detected in at least three channels, \edit{representing 112 separate regions}, 
	and the stars represent averages across frequency \edit{for a given region}.
	\redit{Squares indicate the 5 \% of sources for which the $V/I$ uncertainty is large ($\approx$ 10 \%) 
	due to possible calibration errors that could not be accounted for (see Section~\ref{leakage}).} 
	\textit{Right}: Histogram of $|V|$/$I$, where the white region corresponds to the colored circles 
	and the shaded region corresponds to the stars from the left panel. 
       }}
    \label{fig:ars_hist}
\end{figure*}

Figure~\ref{fig:ars_hist} shows a scatterplot of polarization fraction [$|V|$/$I$] versus the total intensity divided by 
the background level [$I$/$I_{\rm bkgnd}$] in the left panel, along with a simple histogram of $|V|$/$I$ in the right panel. 
These are plotted both for each frequency channel independently and for averages of the same source detected 
in multiple channels. 
The background is defined as the median intensity in pixels detected above 5 $\sigma$, and the noise level 
[$\sigma$] is defined as the standard deviation within a 1-pixel border (1156 pixels) that run along the edge of the
289$\times$289-pixel  ($\pm$ \rsolar{3}) field-of-view. 
We find a very broad range of source intensities, ranging from slightly below the background level 
to 50 times greater, with polarization fractions ranging from a few tenths of a percent to nearly 100 \%. 
\redit{The average source has a Stokes $I$ intensity of 7.6$\times$ the background level and a 
polarization fraction of 27 \%.}
The most striking aspect of Figure~\ref{fig:ars_hist} is the relationship between total intensity \edit{over the background} and polarization 
fraction\edit{, which are positively correlated with a Pearson correlation coefficient [$r$] of 0.64.}


\begin{figure*}
    \centering
    \includegraphics[width=1.0\textwidth]{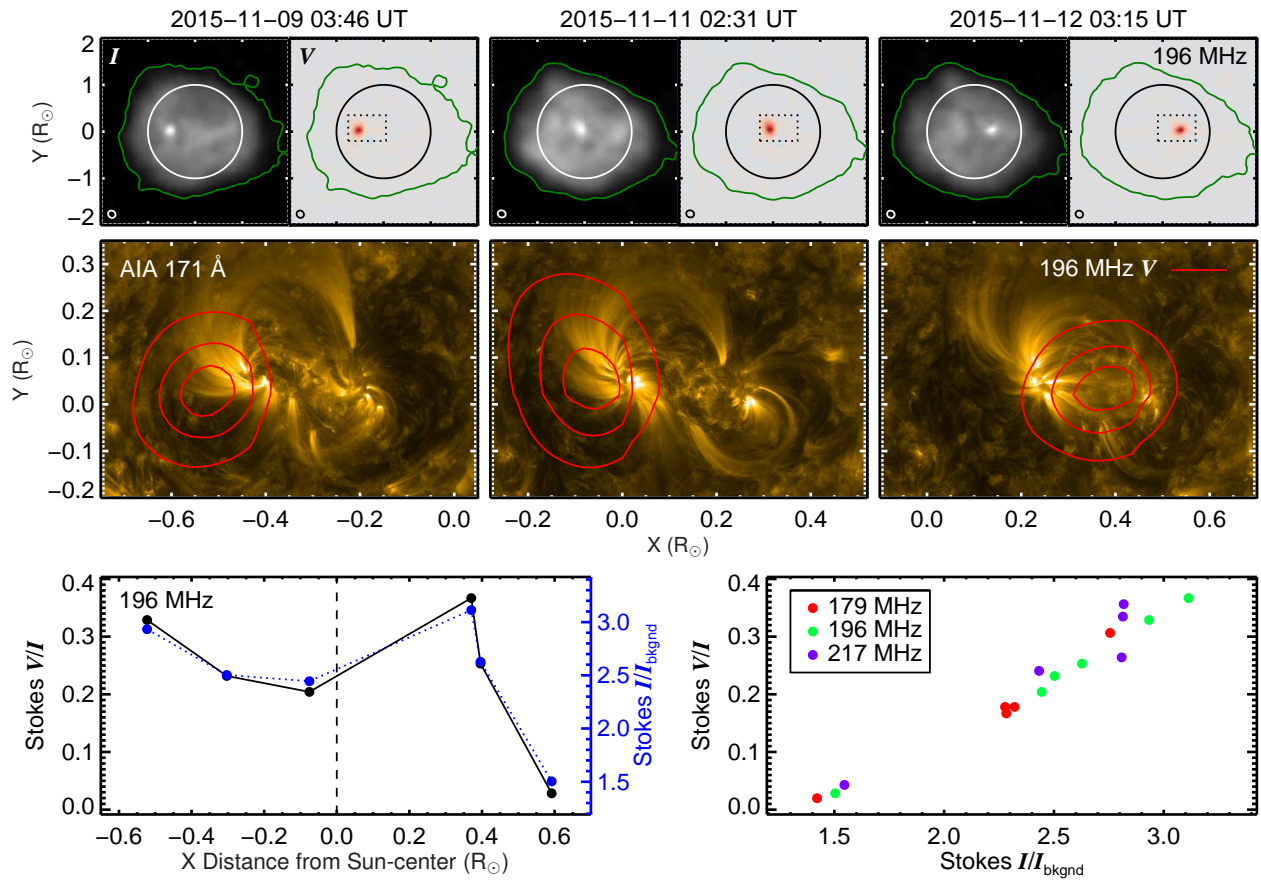}
    \caption{\footnotesize{
	\textit{Top}: The same compact source observed on consecutive days at 196 MHz. 
	\textit{Middle}: Overlays of the 196 MHz circular polarization signal onto 171 \AA{} images from AIA. 
	the field-of-view is marked by the dotted region in the top row, and the contours are at 20, 50, and 80 \% 
	of peak intensity. 
	\textit{Bottom}: The left panel plots polarization fraction [$V$/$I$] and Stokes $I$/$I_{\rm bkgnd}$ 
	as a function of horizontal distance from Sun-center at 196 MHz \redit{for the same source over five days from 9 to 13 November 2015}.
	The right panel plots the same two parameters against each other for three different frequency 
	channels. 
       }}
    \label{fig:ars_lon}
\end{figure*}

Figure~\ref{fig:ars_lon} displays results from one of the \redit{few} sources in our sample for which we have observations 
on several consecutive days and for which \redit{a compact radio source appears in association with the same active region on each day.} 
The source is polarized between 3 \% and 37 \% in the same sense in six observations between \edit{9 and \redit{13} November 2015}. 
Our goal here is to investigate a potential relationship between distance from Sun-center and polarization fraction, 
as previous studies have found noise storms to exhibit higher polarization fractions near disk center. 
We do not find such a relationship \edit{in Figure~\ref{fig:ars_lon}}, and instead this exercise further reinforces the \edit{positive correlation} between 
total intensity and polarization fraction, which apparently becomes stronger if one considers several observations 
of the same source. 
Recall from Section~\ref{mwa} that the observations presented in this paper all represent the baseline intensity over 
5-min observing periods, constructed from the averages of images with the lowest total intensities.
These sources do fluctuate in intensity, so it may be possible to control for the intensity relative to the background 
and then recheck if the polarization fraction has a longitudinal dependence in a future study, ideally with 
more observations of individual sources detected on consecutive days. 

Figure~\ref{fig:ars_lon} also overlays circular polarization contours onto 171 \AA{} images from the 
\textit{Atmospheric Imaging Assembly} (AIA; \citealp{Lemen12}) onboard the \textit{Solar Dynamics Observatory} (SDO; \citealp{Pesnell12}). 
The radio source is associated with a large active region (AR 12448) and is located over fan-loop
structures that represent the bases of loops extending to larger heights. 
This make sense, as the accelerated electrons that are presumably responsible for this radiation must be able to escape to 
sufficiently large heights, corresponding to densities that are sufficiently low for low-frequency emission. 
It is also interesting to note that when the source is east of disk-center, the radio source is associated with the trailing 
sunspot. 
\redit{Noise storms have long been associated with active regions (\textit{e.g.} \citealp{LeSqueren63,Gergely75,Alissandrakis85}), and} 
previous observations have found noise storm sources to be more often associated with the leading spot \citep{White92}. 
A natural followup would be to investigate that aspect systematically for the sources detected here. 
The site of radio emission within the active region also shifts somewhat in time, with the apparent Carrington longitude 
jumping by 12 degrees between \edit{11 and 12 November 2015}. 
This is likely due to evolution in the active region shifting the region where the energetic electrons 
either originate or are able to \edit{reach} in height. 

The main question posed by these results, particularly the scatterplot in Figure~\ref{fig:ars_hist}, is 
whether or not these sources all represent the same basic phenomenon. 
\redit{Our average polarization fraction (27 \%) is lower than previous measurements of noise storm continua. 
Most studies report similar polarization levels for the Type I burst and continuum components of noise storms, which generally 
exceed 80 \% \citep{Elgar77}, but sources with lower polarization fractions have also been reported. 
\citet{Dulk84} observed noise storm continua with polarization fractions of $\approx$ 40 \%, around 15\,--\,20 \% lower 
than the associated bursts. 
As discussed in Section~\ref{leakage}, the most intense and likely highly-polarized sources in our dataset could not be 
included because the leakage mitigation algorithm could not be applied. 
This diminishes the overall average, but more importantly, our sample includes a large number of weak and weakly-polarized sources 
that could not have been characterized by previous instruments.
For example, the 3 September 2015 source ($V/I$ = 3.4 \%) shown in Figure~\ref{fig:ars} is prominent in the polarization map but 
is visually indistinguishable from the quiet Sun in total intensity.}
Given that there does not appear to be any separation into distinct populations \redit{in Figure~\ref{fig:ars_hist}}, we suggest that 
\redit{the data represent} a continuum of plasma emission noise storm \redit{continua} sources with intensities and polarization fractions down to 
levels that were not previously detectable.

\edit{For sources with relatively low total intensities (\textit{e.g.} $I/I_{\rm bkgnd} \lesssim 5$; 61 \% of the population),}
\redit{where the nonthermal component is not entirely dominant,}
very low polarization fractions can be explained by there being a mixture of thermal and nonthermal emission 
within the same resolution element. 
Recall from Section~\ref{introduction} that the thermal bremsstrahlung and nonthermal plasma emission mechanisms 
generally produce opposite polarization signs for the same magnetic field orientation, but the plasma emission 
component is much more highly polarized. 
Therefore, a pixel may be dominated in total intensity by bremsstrahlung emission while the polarized intensity 
is dominated by plasma emission.
The polarization fraction then rises with intensity relative to the background because the relative contribution from 
plasma emission increases. 

Filling factors and beam dilution are also likely to be important, as the thermal component is likely to fill the resolution 
element while the nonthermal component \edit{may come} from a sub-resolution structure. 
This \edit{would mean} that the more highly-polarized nonthermal signal is diluted, which would 
further bring down the polarization fraction. 
\edit{Nonthermal emission sources may not necessarily be intrinsically smaller than the beam size, however.
For instance, \citet{Mohan18} found the scattering-deconvolved sizes of type III burst sources to be significantly larger than the PSF.
As the total intensity becomes much larger than the background and the nonthermal component becomes entirely 
dominant, physical effects related to the emission mechanism and radio wave propagation become increasingly 
important to the interpretation of relatively low polarization fractions in plasma emission sources.
As described in Section~\ref{introduction}, scattering by density inhomogeneities may reduce 
the polarization fraction, as can other propagation effects such as mode coupling.} 
These ideas are discussed further in Section~\ref{discussion} in the context of our other results.


\section{Coronal Holes}
\label{ch}

\begin{figure*}
    \centering
   \includegraphics[width=1.0\textwidth]{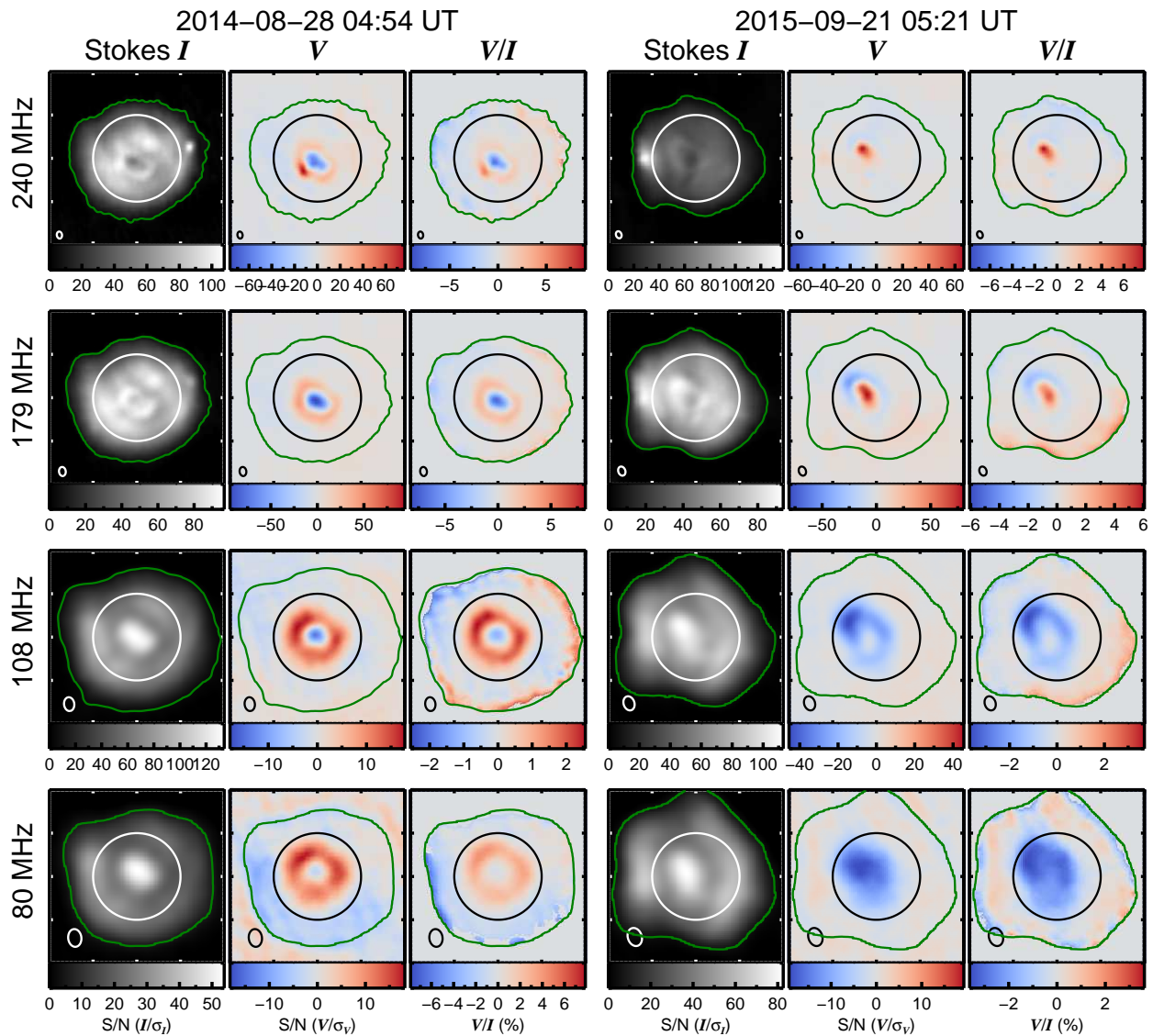}
    \caption{\footnotesize{
	Stokes $I$, $V$, and $V$/$I$ at four frequencies across our bandwidth for two different coronal holes with 
	opposite polarization signatures.
	Color bar units are in signal-to-noise [S/N] for $I$ and $V$ and percent for polarization fraction [$V$/$I$]. 
	The green contours represent the 5 $\sigma$ level in Stokes $I$, the solid circles represent the optical disk, and 
	the ellipses in the lower-left corners represent the synthesized beam sizes. 
       }}
    \label{fig:ch}
\end{figure*}

Perhaps the most surprising finding to immediately emerge from these data is a 
\edit{characteristic ``bullseye"} structure that is frequently exhibited by low-latitude coronal holes and, more 
generally, that coronal holes are the Sun's most prominent features in circular 
polarization at low frequencies in the absence of intense noise storm emission. 
Coronal holes are regions where the magnetic field is open, allowing material 
to freely flow into interplanetary space to form the fast solar wind \citep{Cranmer09}. 
Because the plasma is not confined by closed fields, the densities inside coronal 
holes are considerably lower than in the surrounding corona, and they 
are correspondingly fainter in the soft X-ray and extreme ultraviolet (EUV) observations that are typically 
used to characterize them. 
This is also true at our highest frequencies, which can be seen for two different 
coronal holes in the Stokes $I$ images shown in the upper row of Figure~\ref{fig:ch}. 


As frequency decreases across our bandwidth, many coronal holes transition from 
being relatively dark to relatively bright with respect to their surroundings. 
This effect had been known from a few previous observations \redit{\citep{Dulk74,Lantos87,Lantos99,McCauley17}} and 
was recently characterized in more detail using MWA observations \citep{Rahman19}. 
The mechanism that produces this increase in brightness is unclear, but 
\redit{different authors have suggested} that refraction near the coronal hole boundary   
\redit{may systematically redirect} emission generated outside of the coronal hole to the interior from an 
observer's perspective \redit{\citep{Lantos87,Alissandrakis94,Rahman19}. 
As discussed by \citet{Rahman19},} 
this leads to a ring of enhanced emission around the coronal hole edge, which is apparent 
in our \edit{higher-frequency images} but cannot be distinguished at the lower frequencies, likely due to the 
lower spatial resolution. 
A corollary of this effect is a ring of diminished intensity in the regions from which 
the refracted emission originated, which we see prominently in the low-frequency images 
at the bottom of Figure~\ref{fig:ch}. 


\begin{figure*}
    \centering
    \includegraphics[width=0.8\textwidth]{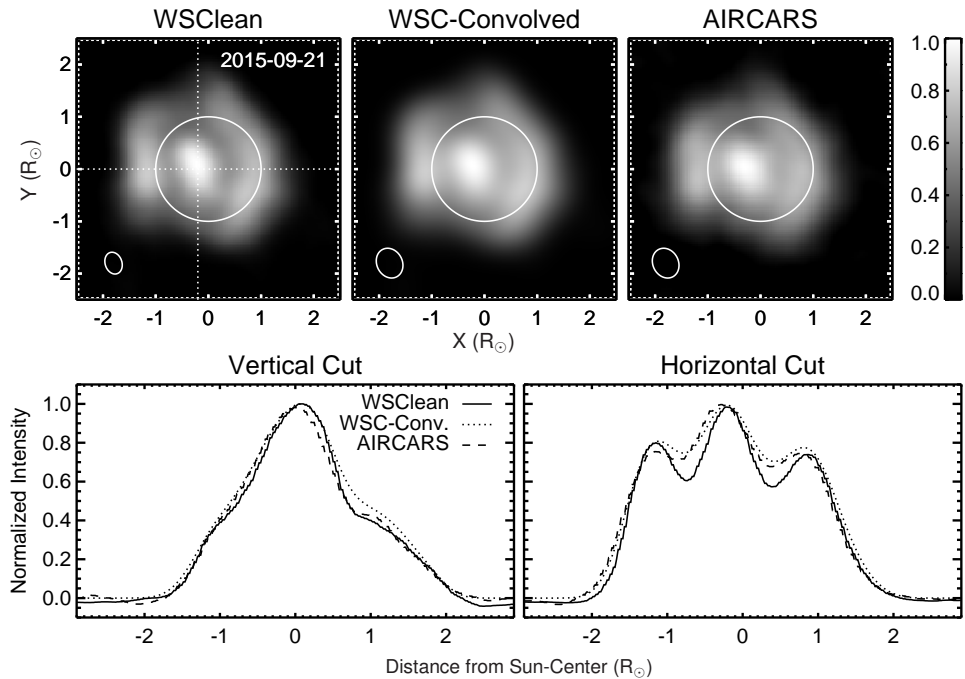}
    \caption{\footnotesize{
	A comparison of the same dataset independently reduced with different calibration techniques 
	and different implementations of the \textsf{CLEAN} algorithm. 
	The Stokes $I$ image at 80 MHz is shown for the \edit{21 September 2015} coronal hole from Figure~\ref{fig:ch}.
	Our reduction on the left, the  \textsf{AIRCARS} reduction on the right, and the  \textsf{AIRCARS} image convolved 
	with the \textsf{WSClean} beam is in the middle. 
	Dotted lines in the left panel indicate the cuts shown in lower two panels. 
	The \textsf{WSClean} and convolved  \textsf{AIRCARS} images appear nearly identical, 
	and the cuts demonstrate that this \edit{is} also true quantitatively. 
       }}
    \label{fig:aircars}
\end{figure*}

The second and third columns of Figure~\ref{fig:ch} show the corresponding 
Stokes $V$ and $V$/$I$ images, respectively. 
Both coronal holes in Figure~\ref{fig:ch} exhibit a central polarized component 
of one sense surrounded by a ring of the opposite sense. 
The outer ring grows in area as the frequency decreases, while the central component 
shrinks until it may or may not be completely gone by 80 MHz. 
This bullseye structure is peculiar \edit{in that coronal holes have unipolar line-of-sight (LOS) 
magnetic field configurations that we expected to result in unipolar Stokes $V$ maps across our 
entire observing band.
We therefore first consider if the feature might be an instrumental or calibration artefact.
While we cannot validate this signature with a completely independent 
observation and data reduction procedure, we have strong evidence to believe that 
this structure is real for the reasons outlined below.}

First, we can validate the structure seen in Stokes $I$ with an independent reduction. 
Figure~\ref{fig:aircars} compares our 80 MHz image of the \edit{21 September 2015} coronal hole, which 
was produced using \textsf{WSClean}, to an image produced using the  \textsf{AIRCARS} pipeline \citep{Mondal19}. 
 \textsf{AIRCARS} uses an entirely different calibration scheme through iterative self-calibration on the Sun itself 
without the need to observe a separate calibrator source. 
This approach is advantageous in that the calibration is tuned to the specific observation of interest, 
which may greatly improve the dynamic range, but it cannot yet be used for polarimetry. 
Our method determines the calibration solutions solely from a known calibrator source, generally 
observed before or after an observing campaign of several hours. 
Figure~\ref{fig:aircars} shows that we obtain nearly identical results from the two pipelines. 
Both methods do use the \textsf{CLEAN} algorithm for deconvolution, as is the standard, 
although implemented through different software packages. 
However, we can be confident that we are not seeing an artefact of the \textsf{CLEAN} because the 
features of interest are also present in the \edit{undeconvolved} ``dirty images". 


\begin{figure*}
    \centering
   \includegraphics[width=1.0\textwidth]{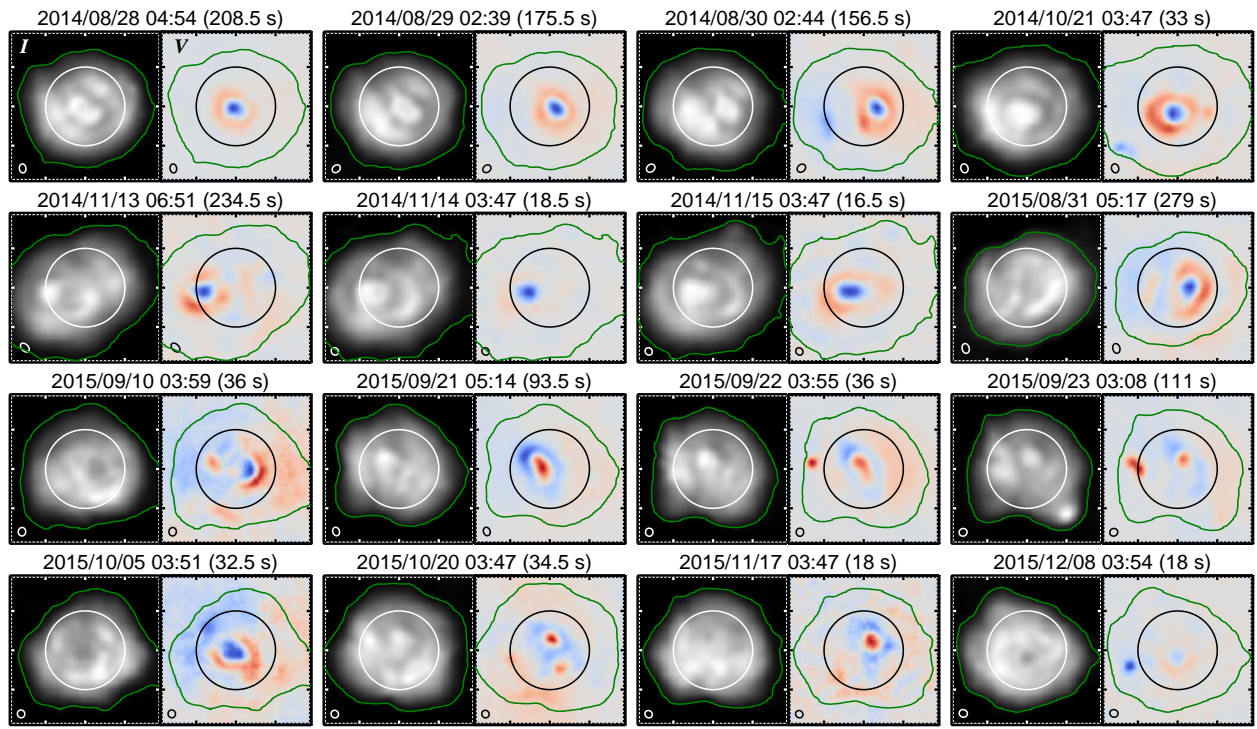}
    \caption{\footnotesize{
	Mosaic of coronal hole observations at 161 MHz, with Stokes $I$ on the left and Stokes $V$ on the right. 
	The green contours show the 5-$\sigma$ levels for Stokes $I$, the solid circles denote the optical 
	limb, and ellipses in the lower-left corners reflect the synthesized beam sizes. 
	The titles correspond to UTC times followed by the effective integration times in parentheses.
	Linear color scales are as in Figure~\ref{fig:ch}, with red, blue, and gray 
	corresponding to positive, negative, and zero Stokes $V$ intensity, respectively.  
	\redit{Images for all 12 frequency channels can been seen in an animated version of 
	this figure available in the \href{http://www.physics.usyd.edu.au/~pmcc8541/mwa/polarization/index.html}{online material}. 
	Note that some dates do not have observations at all frequency channels because of calibration failures, and 
	two observation periods exhibit bright nonthermal sources at higher frequencies that prevent the coronal 
	hole structure from being visible due to limited dynamic range.  }
       }}
    \label{fig:ch_mosaic}
\end{figure*}

\redit{We are also confident that the Stokes $I$ into $V$ leakage subtraction method has not introduced this feature.  
As described in Section~\ref{leakage}, the leakage fraction ($L$) is assumed to be constant across the relatively small 
spatial scale of the Sun based on results from widefield astronomical studies \citep{Lenc17}.
Because $L$ is constant, varying it changes the fractional polarization level without changing the qualitative structure.
Figure~\ref{fig:lsub1} illustrates how the subtraction algorithm works for one of the same coronal hole observations shown in 
Figure~\ref{fig:ch}, and the animated version of Figure~\ref{fig:lsub1} (\href{http://www.physics.usyd.edu.au/~pmcc8541/mwa/polarization/index.html}{available online}) shows how the ``corrected" images look as 
a function of $L$.
The animation shows that the polarization reversal bullseye pattern remains for all values of $L$ until reaching the extremes, 
where 80\,--\,100 \% of pixels across the Sun are too highly polarized of the same sense to be believable. 
Further, varying $L$ by just 1\,--\,2 \% on either side of the value obtained from the correction algorithm quickly pushes 
into this extreme case. 
And even at the extremes, the qualitative ring pattern remains as a sharp change in polarization fraction instead of a reversal.}

\redit{Next}, this polarization ring structure is not rare and seems to be characteristic 
of low-latitude coronal holes across our dataset. 
We have 28 separate observations of 13 different 
coronal holes in 2014 and 2015 that exhibit this effect. 
A mosaic of examples is shown in Figure~\ref{fig:ch_mosaic} at 161 MHz, the center 
of our \edit{observing band}\redit{, and the other channels can be seen in the corresponding animation.}
Several coronal holes in the mosaic are shown on consecutive days, and they move 
with the solar rotation as expected. 
This structure is not observed in association with other solar features, despite 
noise storms often having similar appearances in total intensity at the 
lowest frequencies. 
\edit{Therefore, from both a data reduction perspective and with respect to its association with solar features, 
the bullseye feature does not appear to be consistent with an instrumental effect.}
Moreover, while this feature is a surprise \edit{to us}, we do find various points of 
consistency between the observations and our expectations that will be discussed later. 

\begin{figure*}
    \centering
    \includegraphics[width=1.0\textwidth]{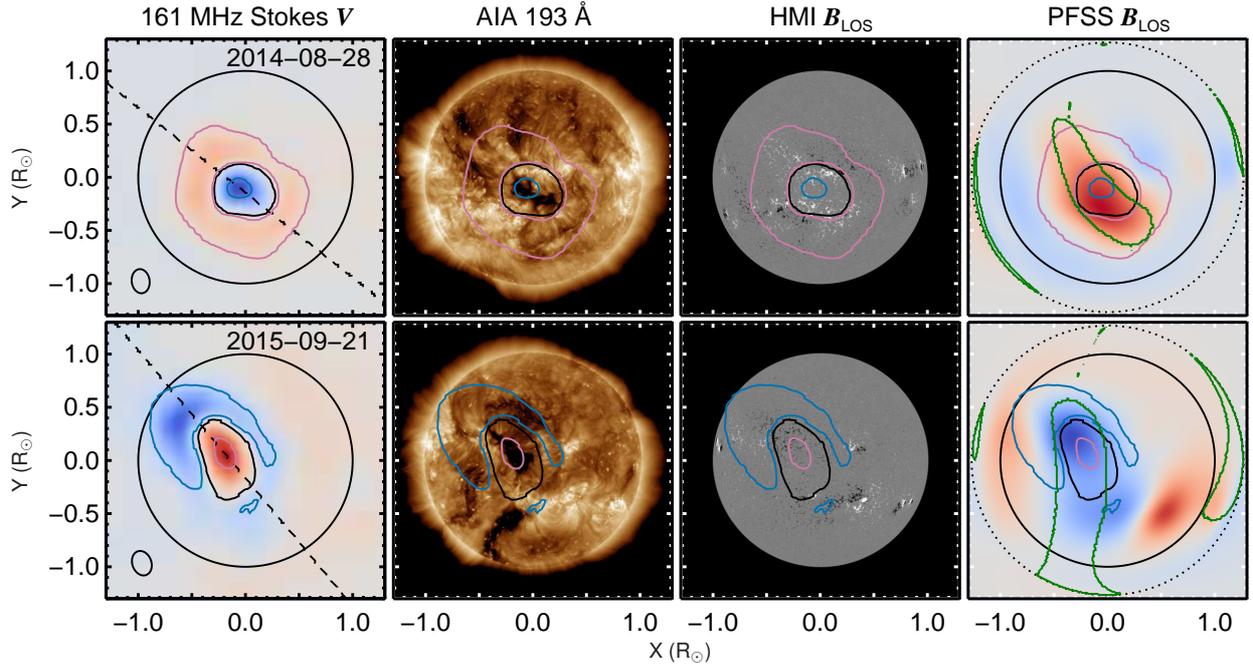}
    \caption{\footnotesize{
	The first column shows Stokes $V$ at 161 MHz for the two coronal holes from Figure~\ref{fig:ch}. 
	The solid black circle represents the optical disk, the black contour is where $V$ = 0 in 
	that region, the blue contour is at 75 \% of the central components' maxima, and the pink 
	contour is at 25 \% of the ring components' maxima. 
	The dashed lines denote the slits used in Figure~\ref{fig:ch_slices}.  
	The second and third columns plot AIA 193 \AA{} and HMI line-of-sight [\redit{$B_{\rm LOS}$}] magnetograms 
	with the $V$ contours from the first column. 
	The last column shows the \redit{$B_{\rm LOS}$} component of the PFSS model 
	at a height roughly corresponding to that of the radio limb (\rsolar{1.27}). 
	Red, blue, and gray colors represent positive, negative, and zero \redit{$B_{\rm LOS}$}, respectively, 
	and green contours represent open field regions in the model at that height. 
       }}
    \label{fig:ch_overlay}
\end{figure*}

Figure~\ref{fig:ch_overlay} overlays contours of the 161 MHz polarimetric signal from the two 
coronal holes shown in Figure~\ref{fig:ch} onto 193 \AA{} images from the 
AIA and LOS magnetograms from the \textit{Helioseismic and Magnetic Imager} (HMI; \citealp{Scherrer12}) 
onboard the SDO.  
These two coronal holes were chosen for this exercise because they have opposite polarization signatures, 
which is consistent with them having opposite magnetic field 
configurations at the photosphere, and because we have two observations of each on the same day, 
which we will use to estimate the uncertainty in $V/I$. 
The last column of Figure~\ref{fig:ch_overlay} overlays the polarization contours onto 
the LOS component [\redit{$B_{\rm LOS}$}] of a potential field source surface (PFSS; \citealp{Schrijver03}) model at a 
height roughly corresponding to that of the radio limb. 
The models were obtained from the PFSS module in SolarSoft IDL and manipulated using the \textsf{FORWARD} codes. 
A height of \rsolar{1.27} is used, which corresponds to the height of the plasma frequency layer at 161 MHz 
in a 3$\times$ \citet{Newkirk61} density model. 


\begin{figure*}
    \centering
    \includegraphics[width=1.0\textwidth]{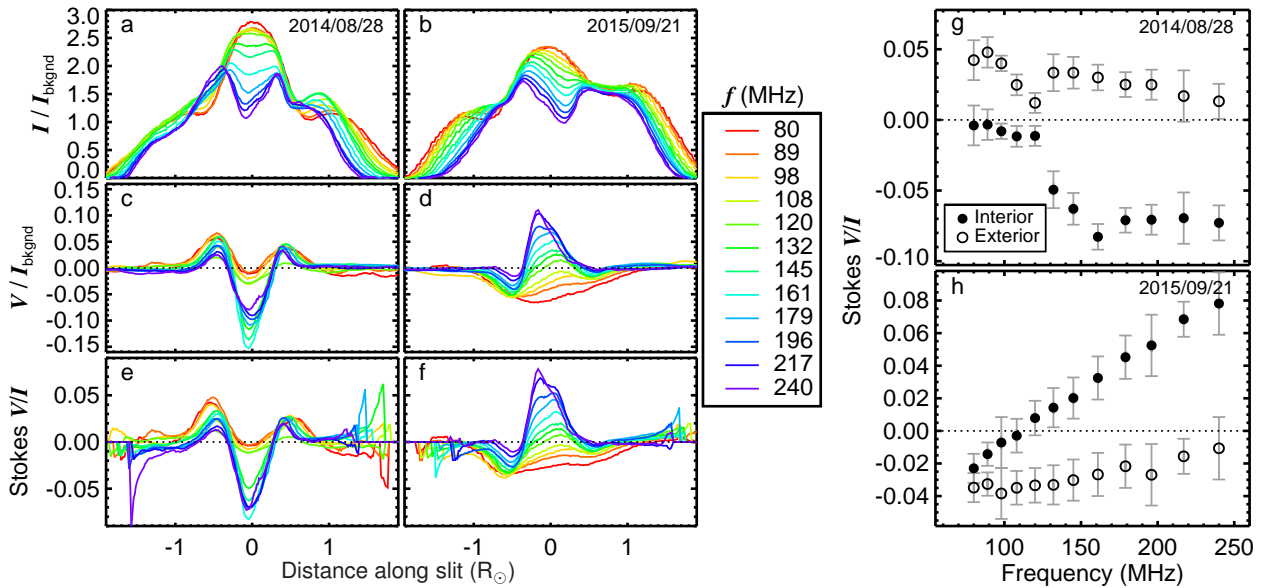}
    \caption{\footnotesize{
	\textbf{(a\,--\,b):} Stokes $I$ intensity for each frequency channel along the slits shown in Figure~\ref{fig:ch_overlay} 
	through two coronal holes on \edit{28 August 2014} (left) and \edit{21 September 2015} (right). 
	\textbf{(c\,--\,d):} Stokes $V$ intensities along the same slits. 
	Panels a\,--\,d are shown in normalized units, where the normalization factor corresponds to the 
	median Stokes $I$ intensity of pixels detected above 5 $\sigma$ (\textit{\textit{i.e.}} the background level). 
	\textbf{(e\,--\,f):} The corresponding Stokes $V/I$ fractional polarization levels.  
	\textbf{(g\,--\,h):} Peak Stokes $V/I$ for the coronal holes' interior (\edit{filled circles}) and exterior ring 
	(open circles) components as a function of frequency. 
       }}
    \label{fig:ch_slices}
\end{figure*}

Crucially, the central polarized components do not match the orientation of \redit{$B_{\rm LOS}$}, as 
would be expected from the thermal bremsstrahlung process that is assumed 
to be the dominant, if not sole, emission mechanism in coronal holes. 
This effect is further characterized in Figure~\ref{fig:ch_slices}, which plots cuts through 
the same two coronal holes in Stokes $I$, $V$, and $V$/$I$. 
The $I$ and $V$ cuts are normalized by the background intensity in Stokes $I$, which we 
define as the median pixel intensity for pixels detected above 5-$\sigma$. 
The gradual transition with decreasing frequency of both coronal holes from being dark to 
bright relative to the background is nicely illustrated by the Stokes $I$ curves for both examples, 
as is the oppositely-oriented ring structure in Stokes $V$. 
Note that the large spikes in $V$/$I$ near the ends of the slits are in locations where both the 
total intensity and polarized signals approach the noise level, making the fractional polarizations unreliable. 

While the overall pattern is similar for both coronal holes, the behavior of the central component is 
somewhat different in each case. 
The rightmost panels (g and h) of Figure~\ref{fig:ch_slices} show the peak $V$/$I$ for the central 
and ring components. 
The central component is most highly polarized ($\approx$ 5\,--\,8\%) at our highest frequency (240 MHz) 
and gradually decreases in polarization fraction with decreasing frequency. 
For the \edit{28 August 2014} example, the central component falls to nearly 0 \% polarization at 80 MHz but remains 
of the same sign at all frequencies, whereas the \edit{21 September 2015} examples crosses 0 \% around 108 MHz and gradually 
approaches the same polarization level as the ring component. 
The latter scenario is somewhat more common in our experience. 
That is, by 80 MHz, the entire source is typically polarized in the same sense expected by bremsstrahlung emission and at a similar level, often with 
a small dip in polarization fraction at the center where the source is oppositely-polarized at higher frequencies.

The uncertainties in panels g and h of Figure~\ref{fig:ch_slices} are the combination of measurement noise 
and \redit{three} effects related to the leakage subtraction algorithm described in Section~\ref{leakage}. 
The first is the range of values found by varying the 
minimization parameter $r_{\rm c,thresh}$ in Equation~\ref{eq:alg1} between 0.3 and 0.8 \%, along with varying the pixels 
included in the operation between those detected above 5 $\sigma$ and those detected above 15 $\sigma$. 
The second \redit{is the difference in} polarization fraction \redit{at} the same \redit{locations} in two observations separated by 2\,--\,3 
hours\redit{, and the third is the potential for unnacounted for leakage of Stokes $V$ into $I$ at up to the same level as that 
measured for $I$ into $V$.}
\redit{These combined uncertainties in $V$/$I$ average $\pm$ 1.2 \% and are as large as $\pm$ 2.9 \%}. 
In both cases, the sign of the leakage fraction flips for each frequency channel between the two observations as the \edit{Sun} moves 
to different locations in the primary beam with respect to the phase center. 
(The MWA does not continuously track an object and instead has a set of discrete pointings that may be changed 
after every $\approx$ 5-min observing period.) 

Despite the sign change in the leakage artefact between observations separated by 2\,--\,3 hours, after implementing the subtraction algorithm, 
the polarization fraction remains consistent to within 1 \% for a given location and frequency channel. 
However, a sharp discontinuity remains between the 120 and 132 MHz channels in the \edit{28 August 2014} observation. 
The leakage is more severe in this observation as compared to the \edit{21 September 2015} data, 
and the sign of the leakage also changes between those two channels. 
The discontinuity \redit{in the polarization fraction trend shown in Figure~\ref{fig:ch_slices}g} 
is therefore likely to be a calibration artefact that cannot be removed by uniformly implementing our correction algorithm. 
\redit{This suggests an additional source of uncertainty in the polarization fraction that is not accounted for by the methods described 
in the previous paragraph. However,}
note that \redit{leakage} affects the polarization level uniformly across the image and cannot warp the qualitative structure observed because the 
leakage does not vary on the small \edit{angular} scale of the Sun, given what we know from widefield astrophysical observations. 

The puzzle with respect to this feature is again the fact that the polarization of the central component does not match the sign expected 
from thermal bremsstrahlung emission. We will discuss possible interpretations for this in Section~\ref{discussion}. 


\section{The Large-Scale Quiescent Structure}
\label{qs}


\begin{figure*}
    \centering
    \includegraphics[width=1.0\textwidth]{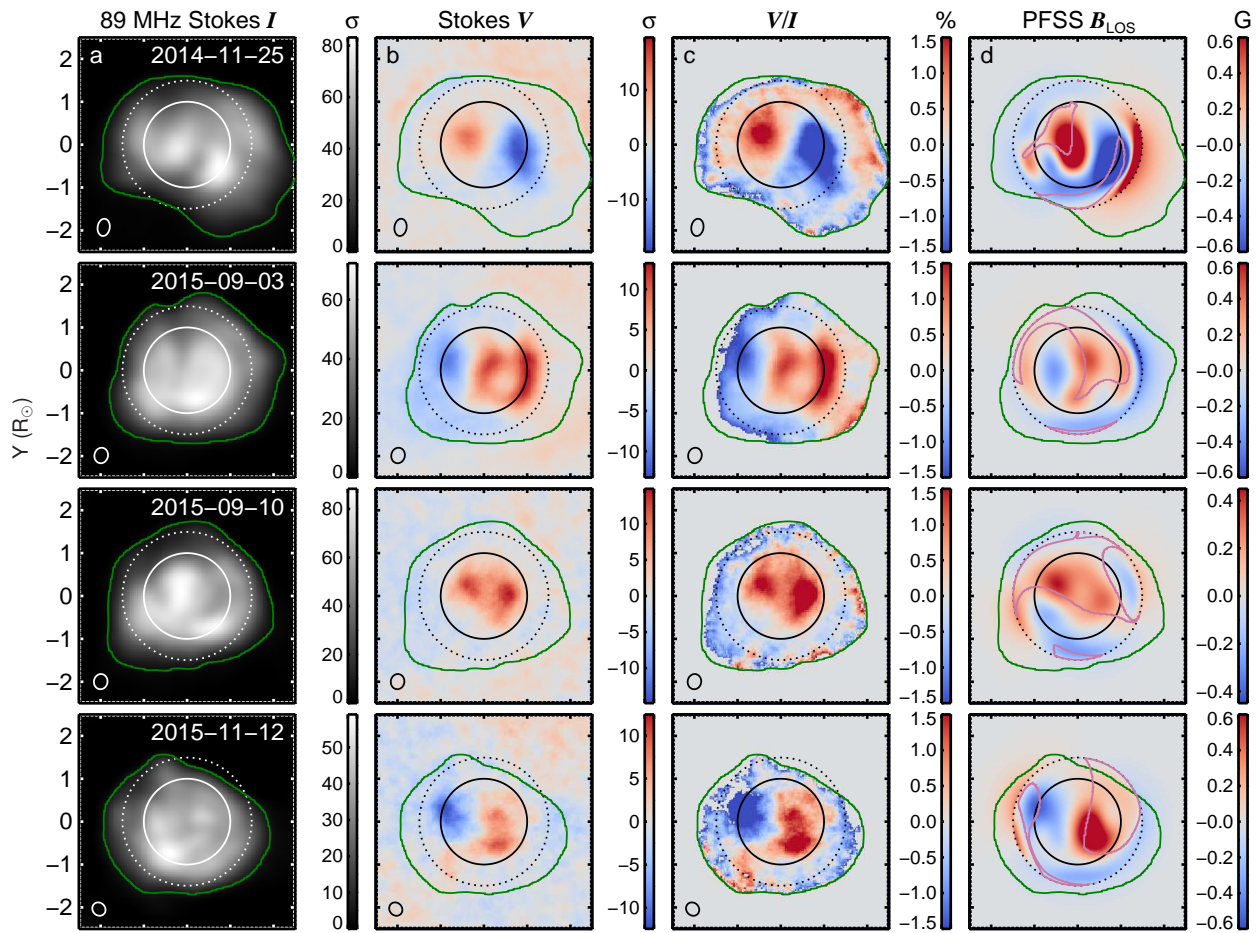}
    \caption{\footnotesize{
	Columns \textbf{a\,--\,c} show Stokes $I$, $V$, and $V/I$ images at 89 MHz on four different days for which 
	a single compact region does not dominate the total intensity. 
	The solid circles denote the optical limb, and the green contours show the 5 $\sigma$ level for Stokes $I$. 
	Ellipses in the lower-left corners reflect the synthesized beams. 
	Column \textbf{d} shows the line-of-sight magnetic field strength [\redit{$B_{\rm LOS}$}] in the PFSS model at a height of \rsolar{1.49}, which 
	roughly corresponds to the height of the radio limb at 89 MHz and is indicated by the dotted circles in each panel. 
	The pink contours indicate open field regions in the model.  
	Color bar units are in signal-to-noise [$\sigma$] for Stokes $I$ and $V$, percent for $V/I$, and Gauss [G] for the field model. 
	\redit{An animated version of this figure that shows all 12 frequency channels is available in the \href{http://www.physics.usyd.edu.au/~pmcc8541/mwa/polarization/index.html}{online material}.}
       }}
    \label{fig:large_low}
\end{figure*}

As mentioned in Section~\ref{introduction}, the thermal bremsstrahlung process that dominates quiescent coronal emission 
at low frequencies produces a slight circular polarization signature in a magnetized plasma that depends primarily on the 
line-of-sight magnetic field strength. 
Absent of other emission mechanisms, a positive LOS field should produce a positive Stokes $V$ signature of up to a few 
percent that depends on the field strength. 

Figure~\ref{fig:large_low} shows Stokes $I$, $V$, and $V/I$ images on four different days for which the polarimetric signature 
is not dominated by a bright noise storm source or disk-center coronal hole. 
These days were also selected to have a mixture of positive and negative Stokes $V$ 
regions so that we can compare the structure to that of the LOS field. 
The fourth column of Figure~\ref{fig:large_low} shows the LOS magnetic field direction 
and strength in the corresponding PFSS model. 
The dotted circle indicates the height at which the model LOS field is shown, which is chosen to be roughly that 
of the radio limb, and the plane-of-sky field is shown beyond the dotted circle. 
This height is \rsolar{1.49} at 89 MHz and corresponds to the height of the plasma 
frequency layer in a 3-fold \citet{Newkirk61} density model. 
Pink contours indicate the open-field regions in the model at the same height, 
which were determined using the ``topology" keyword in the \textsf{FORWARD} code. 
\edit{It is immediately apparent that the Stokes $I$ and $V$ maps 
show very different morphologies in general. 
Regions with the highest polarized intensities are often not straightforwardly correlated with 
those of highest total intensity.}
It is \edit{also} interesting to note that larger polarized intensities are often associated with open 
field regions, which is also consistent with the coronal hole observations from the previous section.  
While we have not investigated this effect systematically, it may be due to there being lower 
densities and lower density contrasts between adjacent regions in open field regions, which 
then reduces the depolarizing effect of scattering by density inhomogeneities. 


\begin{figure*}
    \centering
     \includegraphics[width=1.0\textwidth]{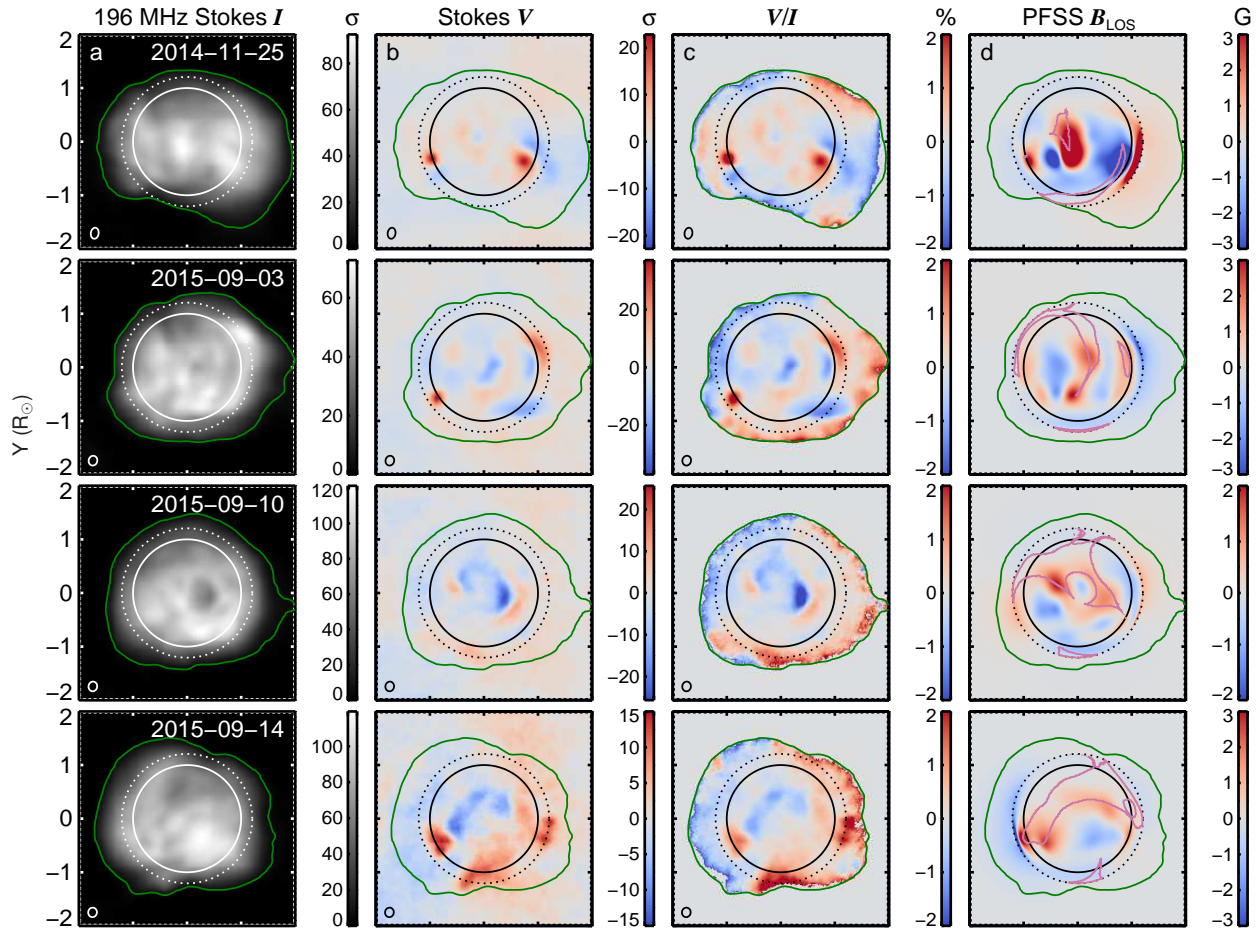}
    \caption{\footnotesize{
	Columns \textbf{a\,--\,c} show Stokes $I$, $V$, and $V/I$ images at 196 MHz on four different days for which 
	a single compact region does not dominate the total intensity. 
	The days are the same as Figure~\ref{fig:large_low} except for the last row. 
	The solid circles denote the optical limb, and the green contours show the 5 $\sigma$ level for Stokes $I$. 
	Ellipses in the lower-left corners reflect the synthesized beams. 
	Column \textbf{d} shows the line-of-sight magnetic field strength [\redit{$B_{\rm LOS}$}] in the PFSS model at a height of \rsolar{1.21}, which 
	roughly corresponds to the height of the radio limb at \edit{196} MHz and is indicated by the dotted circles in each panel. 
	The pink contours indicate open field regions in the model.  
	Color bar units are in signal-to-noise [$\sigma$] for Stokes $I$ and $V$, percent for $V/I$, and Gauss [G] for the field model. 
       }}
    \label{fig:large_high}
\end{figure*}

Figure~\ref{fig:large_low} demonstrates that the Stokes $V$ structure at our lowest frequencies 
is generally well-matched to the LOS field, at least near disk center\edit{, and that the 
sign of Stokes $V$ is broadly consistent with that expected from thermal bremsstrahlung emission given the LOS field orientation}. 
The boundaries between opposite polarization signs are roughly aligned with polarity inversion 
lines in the model. 
The agreement tends to diminish with distance from disk center, which is likely due to two effects. 
First, low-frequency radio emission is heavily influenced by propagation effects, namely refraction, scattering, and mode coupling, 
that can influence the polarization sign and fraction, and these effects 
become more pronounced near the limb \citep{Shibasaki11}. 
Second, although the polarization fraction is expected to be highest off the limb \citep{Sastry09}, the 
intensity is much lower there. 
The third column of Figure~\ref{fig:large_low} shows that we often do find relatively high polarization 
fractions toward the radio limb, but these pixels are very close to the noise level in Stokes $V$ and 
we do not regard them as reliable. 
Missing pixels in the Stokes $V/I$ images inside the green 5 $\sigma$ Stokes $I$ contour 
are censored because they have polarization fractions greater than 1.5 \% but Stokes $V$ 
signals below 5 $\sigma$. 

Figure~\ref{fig:large_high} shows the same parameters as Figure~\ref{fig:large_low} but at 196 MHz. 
The same observation periods are used in the first three rows, but the fourth row is different because 
the \edit{12 November 2015} observation becomes dominated by a noise storm source at higher frequencies. 
\redit{First, we note that the polarized quiet Sun emission is more localized at 196 MHz compared to 89 MHz, 
which is likely the combination of at least three effects. 
First, the spatial resolution is simply lower at lower frequencies. 
Second, lower frequency emission is produced at larger heights where the corona is somewhat less 
finely structured, with smaller contrasts in magnetic field and density between adjacent regions. 
Third, lower-frequency emission is more strongly scattered, 
which leads to increased angular broadening with decreasing frequency.}

\redit{Figure~\ref{fig:large_high} also exhibits} much less straightforward agreement between the LOS field orientation and 
the polarization structure. 
This is a gradual transition with increasing frequency that we cannot attribute to instrumental effects 
because we never observe such an inversion for the noise storm sources that are detected across our 
entire \edit{observing band} from 80 to 240 MHz. 
Some of the differences between the Stokes $V$ sign and that expected from the LOS field orientation 
can likely be attributed to the same unknown effect present in the coronal hole observations from 
Section~\ref{ch}.
For instance in the \edit{10 September 2015} and \edit{14 September 2015} observations, there are coronal holes near the 
west limb and north pole, respectively, that exhibit this effect. 
However, those regions aside, there is still not the same alignment between opposite polarization 
signs and LOS field polarity inversion lines that we see at low frequencies. 

One possible explanation for the discrepancies at higher frequencies is simply the accuracy and resolution 
of the potential field model. 
Higher frequencies correspond to lower heights, and the true coronal magnetic 
field becomes increasingly non-potential closer to the surface, with larger contrasts between adjacent regions. 
This is also true for the density and temperature, which also affect the polarization signal to some extent 
and will be important for future forward modeling efforts. 
However, our impression is that physical effects are also likely to be important, 
and we discuss some possibilities in the next section. 

 
\section{Discussion} %
\label{discussion} %

These data offer an opportunity to probe the coronal magnetic field at heights and 
scales that are not easily accessed with other instruments. 
The intent of this article is to introduce the data and survey the range of features observed 
in this new regime. 
In a forthcoming study, we will directly compare the magnetic field strength and structure 
implied by our observations to model predictions of the thermal bremsstrahlung emission 
implied by different global models.  
Preliminary results suggest that we can successfully generate synthetic Stokes $V/I$ images 
that reproduce the low-frequency polarization structure near disk center, but we have 
yet to explicitly compare the field strengths implied by the observed polarization fractions.

While these preliminary results are encouraging, there are a number of questions 
that will require deeper investigation. 
The most perplexing of these is perhaps the bullseye structure in Stokes $V$ described in 
Section~\ref{ch}, which is found for many low-latitude coronal holes. 
At our higher frequencies ($\leq$ 240 MHz), low-latitude coronal holes often exhibit a central 
circularly-polarized component surrounded by a full or partial ring of the opposite sense, 
and the central component is of the opposite sign than would be expected from thermal 
bremsstrahlung, the presumed emission mechanism. 
With decreasing frequency, the central component diminishes and \edit{the} ring expands. 
By 80 MHz, there may be some trace of the central component remaining or the region may 
be entirely of the sign expected by bremsstrahlung emission. 

We have two suggestions for what may be responsible for this effect. 
The first has to do with differential refraction of the $x$- and $o$-modes. 
The two modes have slightly different group velocities within the plasma and therefore 
slightly different refractive indices, as is often observed in ionospheric propagation experiments at Earth \citep{Melrose86}.
It may be possible to separate the two modes sufficiently via refraction so as to produce a 
polarization sign that is the opposite of that expected by bremsstrahlung emission. 
\edit{For example, the refractive index of the $x$-mode is further below unity than the $o$-mode. 
Emission generated by denser plasma at the edge of a coronal hole may refract at the coronal hole boundary, with 
the $x$-mode refracting more strongly in the radial direction, producing an enhancement of $x$-mode around 
the perimeter and $o$-mode near the center.}
Propagation effects do seem to be particularly important in coronal holes, as they are the 
preferred explanation for why some coronal holes are significantly brighter than their 
surroundings at low frequencies \citep{Alissandrakis94,Rahman19}. 
However, this later question has not been resolved, and it is not obvious to us how 
this polarization signal would be produced. 
Ray tracing simulations, such as those done by \citet{Benkevitch12} or \citet{Vocks18}, 
are likely needed to investigate this further. 

A second possibility that we regard as less likely is that the central polarized component 
is produced by weak plasma emission, which generally produces the opposite sign in circular 
polarization compared bremsstrahlung for the same line-of-sight field. 
Because plasma emission may be up to 100 \% circularly polarized, it is possible 
for the total intensity to be dominated by thermal bremsstrahlung emission while the polarized component 
is dominated by plasma emission. 
This is what we expect to be happening in the low-intensity and weakly-polarized active region sources described 
in Section~\ref{ar}. 
For coronal holes, the driver of this plasma emission might be the transition region network jets 
discovered by the Interface Region Imaging Spectrograph \citep{Tian14}. 
These jets are continuously generated throughout the transition region, but they are 
particularly common and intense inside coronal holes \citep{Narang16}. 
They strongly resemble the coronal X-ray jets associated with Type III bursts, 
and some attribute the network jets to small-scale reconnection events \citep{Tian14,Kayshap18}. 
However, the network jets also resemble chromospheric Type II spicules 
and may instead be driven by shocks \citep{Cranmer15} or heating fronts \citep{Pontieu17}. 
If these jets are associated with reconnection or shocks, then nonthermal electrons capable of producing 
plasma emission may be expected at least to some extent by analogy with solar radio bursts. 
We may see this emission only in coronal holes because the densities there are low 
enough that the plasma levels associated with our highest frequencies are in, 
or very close to, the transition region. 
This idea could be explored observationally through examining variability in the 
polarization signal and comparing that to the typical jet timescales. 

A mixture of plasma emission and thermal emission may also help to explain 
why our higher-frequency Stokes $V$ maps are not well-correlated with the 
LOS field structure. 
One of the MWA's main contributions thus far has been to demonstrate 
the prevalence of very weak nonthermal emissions \citep{Suresh17}. 
\citet{Sharma18} report that up to 45 \% of the total intensity outside of nominal burst periods 
may be nonthermal during moderately active periods, and preliminary imaging analyses  
suggest that nonthermal components are present to varying degrees in every environment. 
These nonthermal emissions are attributed to plasma emission, which again is generally 
much more highly polarized and of the opposite sign compared to bremsstrahlung. 
A relatively minor total intensity contribution from plasma emission would therefore have a 
much greater impact on the circularly-polarized intensity and may be capable of reversing 
the observed sense from that expected from thermal emission. 
The polarization fraction could also potentially be used to disentangle the contributions of 
both mechanisms. 

Mode coupling effects associated with quasi-transverse (QT) regions are also likely to be 
important at least in some regions and may contribute to differences between the \edit{observed} polarization 
sense and that straightforwardly expected from a particular emission mechanism\edit{, along 
with reductions in the polarization fraction expected from plasma emission sources.} 
QT regions refer to when the magnetic field orientation is nearly perpendicular to that of 
the emission region, such that there is no magnetic field component along the ray path. 
Passing through such a region may cause the circular polarization 
sign to reverse if the emission frequency is below a certain threshold that depends on the 
plasma properties \citep{Cohen60,Zheleznyakov64,Melrose94}.
This concept is very important to the interpretation of polarization reversals and associated 
magnetic field diagnostics in high-frequency microwave observations (\textit{e.g.}  \citealp{Ryabov99,Altyntsev17,Shain17,Sharykin18}).
At lower frequencies, QT regions are also invoked to explain the polarization properties of noise storms \citep{Suzuki80,White92}, 
Types U and N bursts \citep{Suzuki80,Kong16}, and zebra patterns in Type IV bursts \citep{Kaneda15}.
A natural place to start investigating the importance of QT regions in our observations would be 
by comparing the polarization sense of the active region noise storm sources from Section~\ref{ar} to 
that expected \edit{from} the magnetic field orientation assuming $o$-mode polarization from plasma emission. 

Lastly, it would be useful to explore improving our calibration approach by imposing 
constraints specific to solar observing. 
We have introduced a strategy, adapted from the astrophysical literature, 
to mitigate an artefact referred to as ``leakage," whereby the polarimetric signal is 
contaminated by some fraction of the total intensity signal. 
This is described in Section~\ref{leakage}. 
While we have demonstrated that our approach is reasonably effective, 
it is clearly not perfect given the discrepancies occasionally observed between adjacent 
frequency channels, as illustrated by Figure~\ref{fig:ch_slices}g. 
A better solution may be available by imposing constraints based on the expectation that  
we can generally assume that the linear polarizations (Stokes $Q$ and $U$) are zero 
because Faraday rotation destroys the linear polarization signal over most observing bandwidths. 
Assuming $Q$ and $U$ are zero implies that the XX and YY instrumental polarizations are 
equal, and this constraint may be applied for each antenna in visibility space, 
allowing for a direction-dependent polarization calibration.
We do not yet know the feasibility of this approach, and new software tools would 
need to be developed to implement it. 

 
\section{Conclusion} %
\label{conclusion} %

We have presented the first spectropolarimetric imaging of the Sun using the MWA. 
These are the first imaging observations of the low-frequency corona that are capable of 
measuring the weak polarization signals outside of intense burst periods. 
We reviewed the two dominant emission mechanisms, thermal bremsstrahlung and plasma 
emission, and how their expected polarization signatures relate to our observations. 
Our data were taken from over 100 observing runs near solar maximum, and we surveyed the 
range of features detected \redit{in quiescent periods}. 
These observations can be used to diagnose the coronal magnetic field at heights and scales 
for which the available data constraints are limited, and this will be the focus of future work. 
Our contributions are as follows: 
\\

\begin{itemize}

\item{We introduced an algorithm to mitigate an instrumental artefact known as ``leakage," whereby some fraction of 
the total intensity [Stokes $I$] signal contaminates the circular polarization [Stokes $V$] images (Section~\ref{leakage}). 
Leakage occurs due to \edit{differences between the actual instrumental response and} 
the primary beam model used to convert images of the instrumental polarizations into 
Stokes \edit{images}. These errors may be due to imperfections in the beam model itself or to other effects that change the instrument's  
effective response, such as individual antenna failures or, importantly, the practice of applying calibration solutions from a calibrator 
source at one pointing to the target source at another pointing. 
We adapted an approach used for astrophysical MWA studies, which show that the leakage varies negligibly over the 
spatial scale of the Sun. 
Given that most of the pixels in our images should be very weakly polarized based on our 
expectations \edit{for} thermal bremsstrahlung emission, we determined the leakage fractions with an algorithm that minimizes the number 
of pixels with polarization fractions [$|V/I|$] greater than 0.5 \%.}
\linebreak

\item{We developed and employed a source finding algorithm that detected around 700 compact sources in the Stokes $V$ images \edit{(Section~\ref{ar})}. 
Only sources detected in a least three frequency channels were analyzed further, corresponding to 112 distinct sources found at 
multiple frequencies. The intensities of these sites ranged from slightly below the background level to 60$\times$ greater than 
the background. Their polarization fractions ranged from less than 0.5 \% to nearly 100 \%. 
\edit{At least one of these sources was present on 78 \% of our observing days, and we found a positive 
correlation between the total intensity over the background and the polarization fraction ($r$ = 0.64).} 
The high-intensity sources with large polarization fractions are \redit{noise storm continua sources} produced by 
plasma emission and associated with active regions. 
As there is no obvious separation of these sources into distinct populations, we suggest that they represent a continuum 
of plasma emission sources down to intensities and polarization fractions that were not previously observable in imaging observations. 
Although the plasma emission theory predicts 100 \% circular polarization for fundamental emission, very low polarizations can be explained 
in this context through three effects. First, \edit{the weaker sources} may still be dominated by thermal bremsstrahlung emission with a 
minor contribution from plasma emission that then dominates the polarized component. 
Second, the plasma emission sites may often be considerably smaller than the beam size, leading to beam dilution that smears the 
polarized signal across a larger area.
Third, scattering by density irregularities may also reduce the polarization fraction\edit{, even for very intense sources.}}
\linebreak

\item{We reported the discovery of a ``bullseye" polarization structure often associated with low-latitude coronal holes in which one polarization 
sense is surrounded by a full or partial ring of the opposite sense \edit{(Section~\ref{ch})}. The polarization of the central component is of the opposite sign 
from that expected from thermal bremsstrahlung, the presumed emission mechanism. Moving from our highest frequency (240 MHz) 
to our lowest (80 MHz), the central component diminishes and the ring expands. Some coronal holes continue to exhibit the ring 
structure at 80 MHz, while others are unipolar in Stokes $V$ with a sense that matches that expected \edit{for} bremsstrahlung emission.
This effect was observed in 28 separate observations of 13 different coronal holes. 
We validated the Stokes $I$ structure with an independent data reduction, and we noted that similar total intensity structures associated 
with noise storms never exhibit this effect. 
We speculated that the structure may be the result of propagation effects, namely refraction, that separate the $x$- and $o$-modes, 
but ray tracing simulations are needed to test this. 
Alternatively, we suggested that the polarization signature may be produced by weak plasma emission produced by the recently-discovered 
transition region network jets that are particularly prevalent inside coronal holes. 
}
\linebreak

\item{We showed that at our lowest frequencies, the large-scale Stokes $V$ structure is reasonably well-correlated with the 
line-of-sight magnetic field structure \edit{obtained from} a global potential field source surface model \edit{(Section~\ref{qs})}. 
The boundaries between opposite polarization signs are generally aligned with polarity inversion lines in the model\edit{, with 
the polarization sign matching the expectation from thermal bremsstrahlung emission given the LOS field orientation.}
The correspondence is best near disk center and diminishes toward the limb, where propagation effects become 
increasingly important and the signal-to-noise decreases. 
At our highest frequencies, there is little straightforward agreement between the LOS field orientation 
and the polarization sign. 
This may be due to the limited accuracy of the potential model, as the coronal field becomes increasingly non-potential 
at lower heights where higher-frequency emission generated. 
However, we suspect that physical effects are also important. 
These may include a mixture of thermal and non-thermal emission in the same region, along with propagation effects such 
as refraction and polarization reversals due to quasi-transverse (QT) regions.
}

\end{itemize}

\begin{acks}  
\edit{This work was primarily supported} by the Australian Government through an Endeavour Postgraduate Scholarship. 
\edit{P. McCauley acknowledges the Asian Office of Aerospace Research and Development (AOARD) of the 
United States Air Force Office of Scientific Research (AFOSR) for travel support through the Windows on Science (WOS)
program.}
\redit{I. Cairns and J. Morgan acknowledge support from AFOSR grants FA9550-18-1-0671 and FA9550-18-1-0473, respectively.}
We thank Don Melrose and Sarah Gibson for helpful discussions. 
\redit{We also thank the anonymous referee for a careful reading and constructive comments.}
This scientific work makes use of the Murchison Radio-astronomy Observatory (MRO), operated by 
the Commonwealth Scientific and Industrial Research Organisation (CSIRO). 
We acknowledge the Wajarri Yamatji people as the traditional owners of the Observatory site. 
Support for the operation of the MWA is provided by the Australian Government's  
National Collaborative Research Infrastructure Strategy (NCRIS), 
under a contract to Curtin University administered by Astronomy Australia Limited. 
We acknowledge the Pawsey Supercomputing Centre, which is supported by the 
Western Australian and Australian Governments.
The SDO is a National Aeronautics and Space Administration (NASA) satellite, and 
we acknowledge the AIA and HMI science teams for providing open 
access to data and software. 
This research has also made use of NASA's Astrophysics Data System (ADS) 
and the Virtual Solar Observatory (VSO, \citealt{Hill09}). 
\end{acks}


  
\bibliographystyle{spr-mp-sola}

\tracingmacros=2
\bibliography{mwa_polarization}  

\IfFileExists{\jobname.bbl}{} {\typeout{}
\typeout{****************************************************}
\typeout{****************************************************}
\typeout{** Please run "bibtex \jobname" to obtain} \typeout{**
the bibliography and then re-run LaTeX} \typeout{** twice to fix
the references !}
\typeout{****************************************************}
\typeout{****************************************************}
\typeout{}}

\end{article} 

\end{document}